\documentclass[published]{JHEP}
\JHEP{10(2001)013}

\usepackage{graphicx}
\usepackage{amsmath}
\usepackage{amssymb}
\usepackage{psfrag}
\usepackage{epsfig}

\newcommand{\UU}{\mathop{\rm {}U}}
\newcommand{\SU}{\mathop{\rm SU}}

\newcommand{\wide}[2]{#2}

\title{Domain walls in QCD} 

\author{Michael McNeil Forbes and Ariel R.~Zhitnitsky\\
  Department of Physics and Astronomy, University of British Columbia\\
  Vancouver, BC, V6T 1Z1, Canada\\
  E-mail: \email{mforbes@physics.ubc.ca}, \email{arz@physics.ubc.ca}}
              
\abstract{QCD was shown to have a nontrivial vacuum structure due to
  the topology of the $\theta\equiv\theta+2\pi n$ parameter.  As a
  result of this nontrivial topology, in the large $N_c$ limit,
  quasi-stable QCD domain walls appear, characterized by a transition
  in the singlet $\eta'$ field.  We discuss the physics of these QCD
  domain walls as well as related axion domain walls and we present a
  new type of axion wall which also contains an $\eta'$ transition.
  We argue that these domain walls are topologically stable in the
  limit $N_c\rightarrow \infty$ and classically stable for large but
  finite $N_c$, however, they can decay through a tunneling process.
  We argue that the qualitative features of these QCD domain walls ---
  namely their classical stability --- persist to the realistic case
  of $N_c=3$ and that it is at least possible that their lifetime
  could be macroscopically large.  If it is, then QCD domain walls
  could play an important role in the evolution of early universe and
  may be detectable in energetic collisions such as those at the
  Relativistic Heavy Ion Collider (RHIC).}

\received{June 12, 2001}
\accepted{October 9, 2001\smash{\llap{\raise-1\baselineskip
      \hbox{{\sc Revised :} October 6, 2001}}}}

\preprint{arXiv:hep-ph/0008315}
\begin{document}

\section{Introduction}

Colour confinement, spontaneous breaking of chiral symmetry, the
$\UU(1)$ problem, $\theta$ dependence, and the classification of
vacuum states are some of the most interesting topics in QCD.
Unfortunately, the progress in our understanding of them is extremely
slow. At the end of the 1970s A.~M.~Polyakov~\cite{Polyakov:1977fu}
demonstrated colour confinement in $3$-dimensional QED (QED$_3$): this
was the first example in which nontrivial dynamics of the ground state
played a key role.  Many papers were written regarding the ground
state structure of gauge theories in the strong coupling regime, but
there were many unanswered questions.  Almost 20 years passed before
the next important piece of the puzzle was
solved~\cite{Seiberg:1994rs,Seiberg:1994aj}. Seiberg and Witten
demonstrated that confinement occurs in supersymmetric (SUSY) QCD$_4$
due to the condensation of monopoles: a similar mechanism was
suggested many years ago by 't Hooft and Mandelstam
(see~\cite{'tHooft:1998pk} for a review).  Furthermore, condensation
of dyons together with oblique confinement for nonzero vacuum angle
$\theta$ was also discovered in SUSY models~\cite{Konishi:1998mk}
(this phenomenon was also argued to take place in ordinary QCD.
See~\cite{'tHooft:1998pk}).

\looseness=-1In addition to providing solid demonstration of earlier ideas, the
recent progress in SUSY models has introduced many new phenomena, such
as the existence of rich vacuum state structure and the existence of
domain walls~\cite{Shifman:1997ua}: topologically stable
interpolations connecting the same vacuum state.  The same conclusion
was reached by Witten in~\cite{Witten:1998xy,Witten:1998uk} based on a
D-brane construction in the limit of large $N_c$. In fact, one can see
that in both approaches, the number of states in the vacuum family is
$N_c$.  Motivated by this development in SUSY gauge theories D-brane
construction, we ask how this applies to QCD.  The issue of
classifying the vacuum states for a given parameter $\theta$ as well
as the phenomenological consequences (domain walls and their decay
etc.\ ) is the subject of this work.  In separate
publications~\cite{Forbes:2000gr} we shall apply these results to
analyze the physics after the QCD phase transition in evolution of
early Universe. In particular, we discuss the possibility of
generation of primordial magnetic fields due to the existence of
short-lived QCD domain walls discussed in this paper.

The starting point of our analysis is an effective Lagrangian
approach.  Experience with SUSY models demonstrates that the effective
Lagrangian approach is a very effective tool for the analysis of large
distance dynamics in the strong coupling regime.  There are two
different definitions of an effective Lagrangian in quantum field
theory.  One of them is the Wilsonian effective Lagrangian describing
the low energy dynamics of the lightest particles in the theory.  In
QCD, this is implemented by effective chiral Lagrangians for the
pseudoscalar mesons.  Another type of effective Lagrangian is defined
by taking the Legendre transform of the generating functional for
connected Green's functions to obtain an effective potential.  This
object is useful in addressing questions about the vacuum structure of
the theory in terms of vacuum expectation values (VEVs) of composite
operators --- these VEVs minimize the effective action.  The latter
approach is well suited for studying the dependence of the vacuum
state on external parameters such as the light quark masses or the
vacuum angle $\theta$.  However, it is not, for example, useful for
studying $S$-matrix elements because the kinetic term cannot be
recovered through this approach.  The utility of the second approach
for gauge theories had been recognized long ago for supersymmetric
models, where the anomalous effective potential was found for both the
pure supersymmetric gluodynamics~\cite{Veneziano:1982ah} and
supersymmetric QCD (SQCD)~\cite{Taylor:1983bp} models.  Properties of
the vacuum structure in the SUSY models were correctly understood only
after analyzing this kind of effective potential.

This paper contains many of the details mentioned in the
letter~\cite{Forbes:2000gr} and is organized as follows:
\begin{description}
\item[Section~\ref{sec:leff}:] Here we review the properties of the
QCD effective Lagrangian~\cite{Halperin:1998rc,Halperin:1998bs} which
is a generalization of the Di Vecchia-Veneziano-Witten (VVW)
Lagrangian~\cite{Witten:1980sp,DiVecchia:1980ve} to include terms
subleading in $1/N_c$ as well as to account for a constraint due to
the quantization topological charge\footnote{Such a generalization was
also motivated by SUSY consideration~\cite{Kovner:1997im}, see
also~\cite{Shifman:1997ua} for a review.}.  One should emphasize from
the very beginning that the specific form of the effective potential
used in this paper is not critical for the present analysis: only the
topological structure and winding $\theta\rightarrow\theta+2\pi n$ ---
a consequence of the topological charge quantization --- is essential.
\item[Section~\ref{sec:topology}:] Here we review the idea of
topological charge conservation and argue that, in the limit
$N_c\rightarrow\infty$, the QCD effective Lagrangian admits stable
domain walls.  We then discuss how these walls are metastable on the
quantum level due to a tunnelling decay mode allowed when we consider
heavier degrees of freedom to the next order in $1/N_c$.
\item[Section~\ref{sec:dwall}:] Here we present the classical domain
wall solutions for QCD domain walls both with and without a dynamical
axion field.
\item[Section~\ref{sec:decay}:] Here we show that the QCD domain walls
are not stable on the quantum level due to a tunnelling phenomena as
discussed in Section~\ref{sec:topology}.  In this section we also
discuss the validity of the large $N_c$ approximation when $N_c=3$ and
argue that the realistic case is qualitatively the same as the large
$N_c$ limit.  We estimate the lifetime of the QCD domain walls when
$N_c=3$ and show that it is potentially macroscopically large.  Thus,
these domain walls may play an important role in various physical
processes.
\item[Section \ref{sec:conclusion}:] This is our conclusion.
\end{description}

\section{Effective Lagrangian and $\theta$ dependence in QCD}
\label{sec:leff}

Our analysis begins with the effective low energy QCD action derived
in~\cite{Halperin:1998rc,Halperin:1998bs}, which allows the
$\theta$-dependence of the ground state to be analyzed. Within this
approach, the pseudo-Goldstone fields and $\eta'$ field are
\pagebreak[1] described by the unitary matrix $U_{ij}$, which
correspond to the $\gamma_{5}$ phases of the chiral condensate: $
\langle \overline{\Psi}_{L}^{i} \Psi_{R}^{j} \rangle = - | \langle
\overline{\Psi}_{L} \Psi_{R} \rangle | \, U_{ij}$ with
\begin{equation}
\label{1}
U = \exp \left[ i \sqrt{2} \, \frac{\pi^{a} \lambda^{a} }{f_{\pi}} + i
\frac{ 2}{ \sqrt{N_{f}} } \frac{ \eta'}{ f_{\eta'}} \right], \quad U
U^{\dagger} = 1\,,
\end{equation}
where $ \lambda^a $ are the Gell-Mann matrices of $ \SU(N_f) $, $
\pi^a $ is the pseudoscalar octet, and $ f_{\pi} = 133$ MeV.  In terms
of $U$ the low-energy effective potential is given
by~\cite{Halperin:1998rc,Halperin:1998bs}:
\wide{
\begin{eqnarray}
&& W_{{\rm QCD}}(\theta,U) =
\label{eq:Leff}\\
&& =-\lim_{V \rightarrow \infty} \frac{1}{V}\,\log
\sum_{l=0}^{N_c-1} \exp \Biggl\{ \frac{V}{2}{\rm Tr} ( M U +
\text{H.c.} )\, + V E \cos \left( \frac{2 \pi l + i\log{\rm Det}~
U-\theta}{N_c} \right) \Biggr\}\,.
\nonumber
\end{eqnarray}
}
{
\begin{multline}
  W_{{\rm QCD}}(\theta,U) = \label{eq:Leff} \\
  -\lim_{V \rightarrow
    \infty} \frac{1}{V}\,\log
  \sum_{l=0}^{N_c-1} \exp \Biggl\{
    \frac{V}{2}{\rm Tr} ( M U + \text{H.c.} )\,
    +V E \cos \left(
      \frac{2 \pi l + i\log{\rm Det}~ U-\theta}{N_c} \right)
  \Biggr\}.
\end{multline}
}
All dimensional parameters in this potential are expressed in terms of
the QCD vacuum condensates, and are well known: $ M = {\rm diag}
(m_{q}^{i}|\langle \bar{\Psi}^{i} \Psi^{i} \rangle|)$; the
constant $E$ is related to the QCD gluon condensate $E = \langle
\frac{b \alpha_s}{32 \pi} G^2 \rangle $, where numerically
$3b=11N_c-2N_f$; the quark condensate $\langle \bar{\Psi}^{i} \Psi^{i}
\rangle \simeq -(240\; {\rm MeV})^3$, and the gluon condensate $\langle
\frac{\alpha_s }{ \pi} G^2 \rangle \simeq 1.2\times 10^{-2}{\rm GeV}^4
$.

It is possible to argue that Equation~(\ref{eq:Leff}) represents the
anomalous effective Lagrangian realizing broken conformal and chiral
symmetries of QCD. The arguments are that Equation~(\ref{eq:Leff}):
\begin{enumerate}
\item correctly reproduces the VVW effective chiral
  Lagrangian,~\cite{Witten:1980sp,DiVecchia:1980ve} in the large $N_c$
  limit, \par [{\small For small values of $ ( \theta - i \log {\rm
      Det} \, U ) $, the term with $ l = 0 $ dominates the infinite
    volume limit.  Expanding the cosine (this corresponds to the
    expansion in $ 1/N_c $), we recover exactly the VVW effective
    potential~\cite{Witten:1980sp,DiVecchia:1980ve} together with the
    constant term $ - E = - \langle b \alpha_s/(32 \pi) G_{\mu\nu}^2
    \rangle $ required by the conformal anomaly:
    \wide{
      \begin{equation}
        \label{3}
        W_{VVW}( \theta, U, U^{\dagger}) = - E - \frac{1}{2} {\rm Tr}(MU +
        \text{H.c.}) + \frac{1}{2} \langle \nu^2 \rangle_{YM} (i \log {\rm
          Det} \, U - \theta )^2\cdots \,,
      \end{equation}}
    {
      \begin{equation}
        \label{3}
        W_{VVW}( \theta, U, U^{\dagger}) = 
        - E  
        - \frac{1}{2} {\rm Tr}(MU + \text{H.c.}) + \frac{1}{2} \langle \nu^2 \rangle_{YM}
        (i \log {\rm Det} \, U - \theta )^2\ldots \;
        ,
      \end{equation}}
    where we used the fact that at large $N_c$, \mbox{$E/N_c^2=-\langle
      \nu^2 \rangle_{YM} $} is the topological susceptibility in pure YM
    theory.  Corrections in $ 1/N_c $ stemming from Equation
    (\ref{eq:Leff}) constitute a new result
    of~\cite{Halperin:1998rc,Halperin:1998bs}.}]
  
\item reproduces the anomalous conformal and chiral Ward identities of
  QCD,\par [{\small Let us check that the anomalous Ward Identities
    (WI's) in QCD are reproduced from Equation~(\ref{eq:Leff}).  The
    anomalous chiral WI's are automatically satisfied with the
    substitution $\theta\rightarrow ( \theta - i \log {\rm Det} \, U
    )$ for any $ N_c $, in accord with
    ~\cite{Witten:1980sp,DiVecchia:1980ve}.  Furthermore, it can be
    seen that the anomalous conformal WI's of~\cite{Novikov:1981xj}
    for zero momentum correlation functions of the operator $
    G_{\mu\nu}^2 $ in the chiral limit $ m_q \rightarrow 0 $ are also
    satisfied when $E$ is chosen as above.  As another important
    example of WI's, the topological susceptibility in QCD near the
    chiral limit will be calculated from Equation~(\ref{eq:Leff}). For
    simplicity, the limit of $ \SU(N_f) $ isospin symmetry with $ N_f
    $ light quarks, $ m_{q} \ll \Lambda_{{\rm QCD}} $ will be
    considered. For the vacuum energy for small $ \theta $ one obtains
    ~\cite{Halperin:1998rc,Halperin:1998bs}
    \begin{equation} 
      \label{4} 
      E_{{\rm vac}} (\theta) = -E + m_q \langle \bar{ \Psi} \Psi \rangle
      N_{f} \cos \left( \frac{\theta}{N_{f}} \right) + {\rm O}(m_{q}^2)\,.
    \end{equation} 
    Differentiating this
    expression twice with respect to $ \theta $ reproduces the chiral Ward
    identities~\cite{Crewther:1977ce,Shifman:1980if}: 
    \wide{
      \begin{eqnarray}
        \lim_{ q \rightarrow 0} \; i \int dx \,
        e^{iqx} \left\langle 0| T \left\{ \frac{\alpha_s}{8 \pi} G \tilde{G} (x) \,
            \frac{\alpha_s}{8 \pi} G \tilde{G} (0) \right\} |0\right\rangle &=& - \frac{
          \partial^{2} E_{{\rm vac}}(\theta)}{ \partial \, \theta^{2}} 
        \nonumber\\
        &=& \frac{1}{N_f} m_q \langle \bar{ \Psi} \Psi \rangle + {\rm O}(m_{q}^2)\,.
        \label{5} 
      \end{eqnarray}
      }{
      \begin{align}
        \label{5}
        \lim_{ q \rightarrow 0} \; 
        i \int dx \, e^{iqx} \langle 0| T \left\{ \tfrac{\alpha_s}{8 \pi} 
          G \tilde{G} (x)  \, 
          \tfrac{\alpha_s}{8 \pi} G \tilde{G} (0) \right\}  |0\rangle &=
        - \frac{ \partial^{2} E_{{\rm vac}}(\theta)}{ \partial \,
          \theta^{2}}
        \\ &= 
        \frac{1}{N_f} m_q \langle \bar{ \Psi} \Psi \rangle  + {\rm
          O}(m_{q}^2) \; .
        \nonumber
      \end{align}
      }
    Other known anomalous WI's of QCD can be reproduced from
    Equation~(\ref{eq:Leff}) in a similar fashion. Consequently,
    Equation~(\ref{eq:Leff}) reproduces the anomalous conformal and
    chiral Ward identities of QCD, and in this sense passes the test
    for it to be the effective anomalous potential for QCD.}]
\item reproduces the known $\theta$
  dependence~\cite{Witten:1980sp,DiVecchia:1980ve}.\\~ [{\small As
    mentioned earlier, our results are similar to those found in
    ~\cite{Witten:1980sp,DiVecchia:1980ve}. A new element which was not
    discussed in 80's, is the procedure of summation over $l$
    in~(\ref{eq:Leff}).  As we shall discuss in a moment, this leads to
    the cusp structure of the effective potential which seems to be an
    unavoidable consequence of the topological charge
    quantization.\footnote{This element was not explicitly imposed in the
      approach of~\cite{Witten:1980sp,DiVecchia:1980ve}: the procedure was
      suggested much later to cure some problems in SUSY models
      (see~\cite{Kovner:1997im} and references therein).  Analogous
      constructions were discussed for gluodynamics and QCD
      in~\cite{Halperin:1998rc,Halperin:1998bs}.}  These singularities are
    analogous to the ones arising in SUSY models and show the
    non-analyticity of phases at certain values of $ \theta $.  The origin
    of this non-analyticity is clear, it appears when the topological
    charge quantization is imposed explicitly at the effective Lagrangian
    level.}]
\end{enumerate}

In general, the $\theta$ dependence appears in the combination
$\theta/N_c$, (see Equation~(\ref{eq:Leff})) which na\"\i{}vely does
not provide the desired $2\pi$ periodicity in the physical
observables.  Equation~(\ref{eq:Leff}), however, explicitly
demonstrates the $2\pi$ periodicity of the partition function.  This
seeming contradiction is resolved by noting that in the thermodynamic
limit, $V \rightarrow \infty$, only the term of lowest energy in the
summation over $l$ is retained for a particular value of $\theta$.
The result is that the local geometry of any particular $\theta$
state, gives the illusion of $\theta/N_c$ periodicity in the
observables, but when one considers the full topology of all the
$\theta$ states and properly switches to the lowest energy branch, one
regains the true $\theta$ periodicity. Of course, the values $\theta$
and $\theta + 2 \pi n$ are physically equivalent for the entire set of
states, but relative transitions --- switching branches --- between
different $\theta$ states have physical significance. It is exactly
these transitions --- resulting from the non-local effects of the
topology of the fields --- that are responsible for the domain walls
we discuss in this paper.

When considering only the lightest degrees of freedom, as we do in the
thermodynamic limit $V \rightarrow \infty$ of~(\ref{eq:Leff}), the
effective potential acquires a cusp singularity where one switches
from one branch of the potential to another.  These cusps represent
physical transitions in heavier degrees of freedom which have been
integrated out to get the low-energy effective
potential~(\ref{eq:Leff}).  The physical effects of these heavier
degrees of freedom will be discussed in
Sections~\ref{sec:heavy-degr-freed} and~\ref{sec:cusps}.  The reader
is referred to the original
papers~\cite{Halperin:1998rc,Halperin:1998bs} for more detailed
discussions of the properties of the effective
potential~(\ref{eq:Leff}).

Our final remark regarding~(\ref{eq:Leff}).  The appearance of the cosine
interaction, $\cos(\theta/N_c)$, implies\footnote{As we noticed in the
  introduction, the specific form of the potential is not very
  essential for what follows.  However, this form is very appealing
  for the present study because with it, we can describe some of the
  domain walls in the analytical form (see below, for example
  (\ref{eq:QCDetasol}).} the following scenario in pure gluodynamics
($\phi_i$'s frozen): the $(2k)^{\rm th}$ derivative of the vacuum
energy with respect to $\theta$, as $\theta\rightarrow 0$, is
expressed solely in terms of one parameter, $1/N_c$, for
arbitrary $k$:
\wide{
  \begin{equation}
    \label{6}
    \left.\frac{ \partial^{2k} E_{{\rm vac}}(\theta)}{ \partial \,
        \theta^{2k}}\right|_{\theta=0} \sim \int \prod_{i=1}^{2k} d^4x_i
    \left\langle Q(x_1)\cdots Q(x_{2k})\right\rangle \sim
    \left(\frac{i}{N_c}\right)^{2k}\,,
  \end{equation}}
{
\begin{equation}
  \label{6}
  \left.\frac{ \partial^{2k} E_{{\rm vac}}(\theta)}{ \partial \,
      \theta^{2k}}\right|_{\theta=0}
  \sim \int \prod_{i=1}^{2k} d^4x_i \left\langle  
    Q(x_1)...Q(x_{2k})\right\rangle
  \sim \left(\frac{i}{N_c}\right)^{2k},
\end{equation}
}
where, $ Q=\frac{\alpha_s}{8 \pi} G \tilde{G} $. This property was
seen as a consequence of Veneziano's solution of the $\UU(1)$
problem~\cite{Veneziano:1979ec}. The reason that only one factor
appears in Veneziano's calculation is that the corresponding
correlation function, $\sim \int \prod_{i=1}^{2k} d^4x_i \langle
Q(x_1)\cdots $ $Q(x_{2k})\rangle $, becomes saturated at large distances
by the Veneziano ghosts whose contributions factorize exactly, and was
subsequently interpreted as a manifestation of the $\theta/N_c$
dependence in gluodynamics at small $\theta$.  However, at that time
it was incorrectly assumed that such a dependence indicates that the
periodicity in $\theta$ is proportional to $2\pi N_c$. We now know
that the standard $2\pi$ periodicity in gluodynamics is restored by
the summation over $l$ in~(\ref{eq:Leff}) such that one jumps from one
branch to another at $\theta=\pi$.

\subsection{Vacuum states}

In the next section we shall discuss different types of domain walls
which interpolate between various vacuum states, but first we should
study the classification of vacuum states themselves. In order to do
so, it is convenient to parameterize the fields $U$ as
\begin{equation}
U=\left(\begin{array}{cccc}
e^{i\phi_{1}}& 0 & \cdots & 0\\
0 & e^{i\phi_{2}}& \cdots & 0\\
\vdots & \vdots & \ddots & 0\\
0 & 0 & 0 & e^{i\phi_{N_f}}
\end{array}
\right)
\end{equation}
such that the potential~(\ref{eq:Leff}) takes the form
\begin{equation}
\label{7}
V(\phi_i,\theta) = - E \cos \left( \frac{1}{N_c} \theta -
\frac{1}{N_c} \sum \phi_{i} \right) - \sum M_{i} \cos \phi_i\,.
\end{equation}
The minimum of this potential is determined by the following equation:
\begin{equation}
\label{8} 
\frac{1}{N_c}\sin \left( \frac{1}{N_c} \theta - \frac{1}{N_c} \sum
\phi_i \right) = \frac{M_i}{E} \, \sin \phi_i \,, \qquad i = 1,
\ldots, N_f \,.
\end{equation} 
At lowest order in $ 1/N_c $ this equation coincides with that
of~\cite{Witten:1980sp,DiVecchia:1980ve}. For general values of $
M_{i} / E $, it is not possible to solve Equation~(\ref{8})
analytically, however, in the realistic case $ \varepsilon_{u},
\varepsilon_{d} \ll 1 \, , \, \varepsilon_s \sim 1 $ where $
\varepsilon_{i} = N_cM_{i}/E$, the approximate solution can be found:
\begin{eqnarray}
  \sin \phi_{u} &=& \frac{ m_d \sin \theta }{ [m_{u}^2 + m_{d}^2 + 2
    m_{u} m_{d} \cos \theta ]^{1/2} } + {\rm O}(\varepsilon_{u},
  \varepsilon_{d}) \,,
  \nonumber \\
  \sin \phi_{d} &=& \frac{ m_u \sin \theta }{ [m_{u}^2 + m_{d}^2 + 2
    m_{u} m_{d} \cos \theta ]^{1/2} } + {\rm O}(\varepsilon_{u},
  \varepsilon_{d})\,,
  \nonumber \\
  \sin \phi_{s}&=&  {\rm O}(\varepsilon_{u},\varepsilon_{d}) \,.
  \label{9}
\end{eqnarray}
This solution coincides with the one
of~\cite{Witten:1980sp,DiVecchia:1980ve} to leading order in $
\varepsilon_{u},\varepsilon_{d} $.  In what follows for the numerical
estimates and for simplicity we shall use the $\SU(2)$ limit $m_u=m_d
\ll m_s$ where the solution~(\ref{9}) can be approximated as:
\wide{
  \begin{equation}
    \begin{array}{rclcrclcrclcrcl}
      \phi_{u} &\simeq& {\theta}/{2}\,, &\qquad& \phi_{d} &\simeq&
      {\theta}/{2}\,, &\qquad& \phi_{s} &\simeq& 0\,, &\qquad& 0 &\leq
      &\;\theta < \pi\,,
      \\
      \phi_{u} &\simeq& ({\theta+2\pi})/{2}\,,&\qquad&
      \phi_{d}  &\simeq& ({\theta-2\pi})/{2}\,, &\qquad&
      \phi_{s} &\simeq& 0\,, &\qquad&
      \pi&\leq & \;\theta < 2\pi\,,\qquad \text{etc}\,.
    \end{array}
    \label{10}
  \end{equation}
  }{
  \begin{align}
    \label{10}
    \phi_{u} &\simeq  \theta/2 ,&
    \phi_{d} &\simeq  \theta/2 ,&
    \phi_{s} &\simeq 0 ,&
    0\leq   &\;\theta < \pi,\\
    \phi_{u} &\simeq (\theta+2\pi)/2 ,&
    \phi_{d}  &\simeq (\theta-2\pi)/2 ,&
    \phi_{s} &\simeq 0 , &
    \pi\leq & \;\theta < 2\pi, & \text{etc}. \nonumber
  \end{align}}
Once solution~(\ref{10}) is known, one can calculate the vacuum energy
and topological charge density \mbox{$Q=\langle 0 |
  \tfrac{\alpha_{s}}{8 \pi } G \tilde{G} |0 \rangle$} as a function of
$\theta$. In the limit \mbox{$m_u=m_d\equiv m$}, \mbox{$\langle
  \bar{d} d \rangle = \langle \bar{u} u \rangle \equiv \langle
  \bar{\Psi} \Psi \rangle $} one has:
\begin{eqnarray}
V_{{\rm vac}}(\theta)&\simeq& V_{\theta=0}+2m\big|\!\langle \bar{\Psi} \Psi \rangle \!\big|
\left(1-\bigl|\cos\frac{\theta}{2}\bigr|\right) 
\nonumber \\
\langle \theta | \tfrac{\alpha_{s}}{8 \pi } G \tilde{G} | \theta
\rangle &=&-\frac{\partial V_{{\rm vac}}(\theta)}{\partial\theta}=
-m\big|\!\langle 0|\bar{\Psi} \Psi |0
\rangle\!\big|\sin\frac{\theta}{2}\,.
\label{11}
\end{eqnarray}
As expected, the $\theta$ dependence appears only in combination with
$m$ and goes away in the chiral limit.  One can also calculate the
chiral condensate $ \langle \bar{\Psi}^{i}_L \Psi^{i}_R \rangle $ in
the $\theta$ vacua using solution (\ref{10}) for vacuum phases:
\begin{eqnarray}
\langle \theta |\bar{\Psi} \Psi |\theta \rangle
&=&\cos\frac{\theta}{2} \;\langle 0|\bar{\Psi} \Psi
|0\rangle_{\theta=0}\,,
\nonumber\\
\langle \theta |\bar{\Psi}i\gamma_5\Psi |\theta\rangle 
&=& -\sin\frac{\theta}{2} \; \langle 0|\bar{\Psi} \Psi|0 \rangle_{\theta=0}\,.
\label{12}
\end{eqnarray}

A remark is in order.  As is well known, in thermal equilibrium and in
the limit of infinite volume, the $|\theta\rangle$ vacuum state is a
stable state for all values of $\theta$.  Thus, it is possible to
conceive of a world with ground state $|\theta\rangle$ where $\theta
\neq 0$.  The physics of this world would be quite different from that
of our own: In particular, P and CP symmetries would be strongly
violated due to the non-zero value of the P and CP violating
condensates~(\ref{11}), (\ref{12}).  Despite the fact that the state
$|\theta\rangle$ has a higher energy than $|0\rangle$~(\ref{11}), it
is stable because of a superselection rule: There exists no gauge
invariant observable $\cal{A}$ in QCD that can communicate between
different $|\theta\rangle$ states, i.e.\
$\langle\theta'|\cal{A}|\theta\rangle\sim \delta(\theta-\theta')$ for
all gauge invariant observables $\cal{A}$.  Therefore, there are no
possible transitions between these
states~\cite{Jackiw:1976pf,Callan:1976je,Callan:1978gz} and any such
state $|\theta\rangle$ may happen to be the ground state for our
world.

On account of this superselection rule, one might ask why $\theta=0$
is so finely tuned in our universe.  Indeed, within standard QCD,
there is no reason to prefer any particular value of $\theta$.  This
is known as the strong CP problem.  One of the best solutions to this
problem has been known for many years: introduce a spontaneously
broken symmetry (Peccei-Quinn symmetry~\cite{Peccei:1977hh}).  The
corresponding pseudo-Goldstone particle --- the axion
\cite{Shifman:1980if}, \cite{Weinberg:1978ma}--\cite{Peccei:1989} ---
behaves exactly like the parameter $\theta$ but is now a dynamical
field, thus we can absorb the parameter $\theta$ by redefining the
axion field $a(x)$ and set $\theta=0$.  The axion is now dynamical and
so the corresponding states $|\theta\rangle \sim |a(x)\rangle$ are no
longer stable: the axion field relaxes to the true minimum $\theta
\sim a(x) = 0$~(\ref{11}).  The axion is included in the
potential~(\ref{eq:Leff}) through the Yukawa interaction
\begin{equation}
\frac{1}{2}V{\rm Tr}(MU + \text{H.c.}) \rightarrow \frac{1}{2}V{\rm
Tr}(MUe^{ia} + \text{H.c.})
\end{equation}
and kinetic term $f_a^2(\partial_\mu a)^2/4$.  Examining the
potential~(\ref{eq:Leff}) we see that the parameter $\theta$ can be
absorbed into the $\UU(1)$ phase of $U$ which in turn can be removed
by a redefinition of the axion field $a\rightarrow a+ \theta/N_f$.
Although axions have not been detected and experiments have ruled out
the possibility of the original electroweak scale
axion~\cite{Weinberg:1978ma,Wilczek:1978pj}, there is still an allowed
window with very small coupling constant $f_\pi/f_a \ll 1$ emphasizing
that the axion arises from a very different scale than the electroweak
or QCD.  Axions with this scale, however, are strong dark matter
candidates (see for example~\cite{Kim:1987ax}--\cite{Turner:1999xj}).
The axion thus provides a way for the vacuum state to relax to the
lowest energy state $|\theta=0\rangle$~(\ref{11}).  In the following
we shall consider two types of domain walls: Axion domain walls where
$\theta$ is the dynamical axion field $a$ and QCD domain walls where
$\theta$ is the fundamental parameter of the theory.  In the latter
case, we assume that the strong CP problem has been solved by some
means and take $\theta=0$ to be fixed.

The effective potential~(\ref{eq:Leff}) can be used to study the
vacuum ground state~(\ref{11}), (\ref{12}) as well as the
pseudo-Goldstone bosons as its lowest energy excitations.  In
particular, one could study the spectrum as well as mixing angles of
the pseudo-Goldstone bosons by analyzing the quadratic fluctuations in
the background field (\ref{11}), (\ref{12}). We refer to the original
papers~\cite{Halperin:1998rc,Halperin:1998bs,Fugleberg:1998kk} on the
subject for details, but here we want to quote the following mass
relationships for the $\eta'$ meson to be used in the following
discussions:
\begin{equation}
\label{13}
f_{\pi}^2m_{\eta'}^2=\frac{4N_f}{N_c^2}E+\frac{4}{N_f}\sum_{i}
m_i\bigl|\langle 0|\bar{\Psi}^i \Psi^i|0 \rangle\bigr| +{\rm O}(m_q^2)\,,
\end{equation}
where $i$ runs over the three flavours $\{u,d,s\}$. This relation is
in a good agreement (on the level of 20$\%$) with phenomenology.  In
the chiral limit this formula takes especially simple form
\begin{equation}
\label{14}
m_{\eta'}^2=\frac{4N_f}{f_{\pi}^2N_c^2}E \,,
\end{equation}
which demonstrates that, in the chiral limit, the $\eta'$ mass is
proportional to the gluon condensate, and is therefore related to the
conformal anomaly.  What is more important for us in this paper is
that the combination on the right hand side of this equation exactly
coincides with a combination describing the width of the QCD domain
wall (see Equation~(\ref{eq:QCDwidth})).  For this reason, the
properties of the QCD domain walls are dominated by the $\eta'$ field.

\subsection{Large $N_c$ limit}

At this point, we pause to consider the large $N_c$ limit first
discussed by 't Hooft~\cite{'tHooft:1974jz}. In order to define the
limit, we must hold $g^2 N_c\sim 1$.  From this it follows that
$\alpha_s\sim g^2\sim N_c^{-1}$.  Similarly, by analyzing the
structure of the appropriate correlation functions, we find the
following leading $N_c$ dependences:
\begin{equation}
\begin{array}{rclcrclcrcl}
\alpha_s &\sim& N_c^{-1}\,, &\qquad& b &\sim& N_c\,, &\qquad& \langle
GG \rangle &\sim& N_c^2\,,
\\
E &\sim& N_c^2\,, &\qquad& f_\pi^2 &\sim& N_c\,, &\qquad&
\langle\bar{\Psi}\Psi\rangle &\sim& N_c\,.
\end{array}
\end{equation}
Applying these to~(\ref{14}) we reproduce Witten's famous
result~\cite{Witten:1979kh}
\begin{equation}
\label{eq:metaNc}
\lim_{m_q\rightarrow 0} m_{\eta'} \sim N_c^{-1/2}\,.
\end{equation}
Here we have first taken the chiral limit $m_q \rightarrow 0$ and then
the large $N_c$ limit.  In general, when $m_q\neq 0$, there are O$(1)$
corrections arising from the second term in~(\ref{13}), but in
reality, the chiral limit is good due to the small quark masses and we
wish to preserve the qualitative behaviour of this limit as we take
$N_c$ to be large.  Thus, in the following discussion, we always take
the chiral limit first so that, to leading order in $m_q$ and $N_c$,
the first term of~(\ref{13}) dominates and~(\ref{eq:metaNc}) holds.

Most importantly, we see that, in the large $N_c$ limit, the $\eta'$
becomes light, and thus we are justified in including its dynamics in
the low-energy effective theory~(\ref{eq:Leff}).  We shall return to
this issue later in Section~\ref{sec:large-n_c-limit} where we will
use the large $N_c$ limit to regain theoretical control.

\section{Topological stability and instabilities}
\label{sec:topology}

The domain walls that we will discuss are examples of topological
defects: classical solutions to the equations of motion which are
stable due to the topological configurations of the fields.  Examples
of topological configurations abound in the literature, for example,
Instantons, Skyrmions, Strings, Domain walls
etc.~\cite{Rajaraman:1982,Vilenkin:1994}

The basic idea is that the theory contains some conserved charge
density $q_\mu$ that is a total derivative $q_\mu=\partial_\mu f(x)$.
In this case, the total charge $Q=\int\!q_0\:{\rm d}^3 x$ is quantized
and represents some topological property of the fields such as winding
or linking.  Magnetic charge in the Georgi-Glashow model is an example
(see~\cite{Rajaraman:1982} for details).

The essential point is that the charges $Q$ are exactly conserved
quantities: the subspace of configurations with $Q=Q_1$ is orthogonal
to the subspace with $Q=Q_2\neq Q_1$.  In particular, $Q_2-Q_1 = n$ so
there is no continuous way to vary a configuration from one subspace
to the other and thus there is no overlap.  Thus, objects with
non-zero charges are absolutely stable: even if have an energy higher
than the true vacuum state where $Q=0$.

Referring back to our effective theory, we shall show that there is a
conserved topological charge associated with domain wall
configurations.  Physically, the states $\theta=0$ and $\theta= 2\pi
n$ are identical: they represent the \emph{same} state.  However, as
one can see from Equation~(\ref{10}), the solutions for the ground
states corresponding to $\theta=0$ and $\theta= 2\pi$ are not
described in the same way: $\theta=0$ corresponds to
$\phi_{u}=\phi_{d}=0$ while $\theta=2\pi$ corresponds to $\phi_{u}=
2\pi,~ \phi_{d}=0$.  It clear that the physics in both these states is
exactly the same: If we lived in one of these state and ignored the
others, then we could assign an arbitrary phases $\sim 2\pi n$ for
each $\phi_{u}$ or $ \phi_{d}$ separately and independently.  However,
if we want to interpolate between these states to get feeling about
both of them, the difference in phases between these states can no
longer be a matter of choice, but rather is specified by
Equation~(\ref{10}).  This classification arises because the singlet
combination $\phi_S=\sum\phi_i$ really lives on a $\UU(1)$ manifold
which has the topology of a circle.  The integer $n$ is only important
if we are discussing transitions around the $\UU(1)$ circle: in this
case, it is important to keep track of how many times the field winds
around the centre.  Thus, $n$ is a topological winding number which
plays an important role when the physics can interpolate around the
entire $\UU(1)$ manifold. We illustrate this idea in
figure~\ref{fig:2dhomo}.  Here, we show three topologically distinct
paths in a two-dimensional space with an impenetrable barrier in the
centre.  The paths that wind around the barrier cannot be deformed
into the other paths. Each path is characterized by a winding number
$n$.
 
\FIGURE[t]{\centerline{\epsfig{file=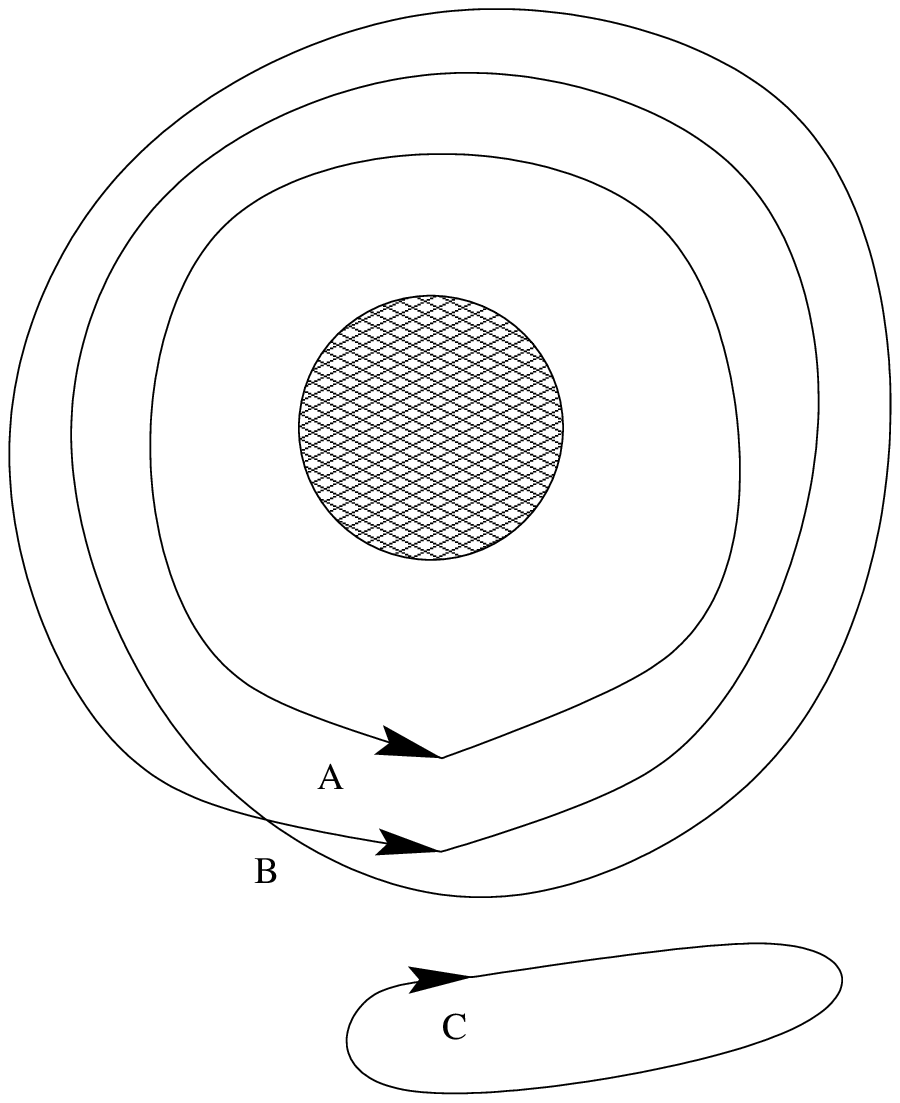,width=0.50\textwidth}}
\caption{Three examples of transitions that are topologically
different.  Paths $A$ and $B$ wind around the hold once ($n=1$) and
twice ($n=2$) respectively whereas path $C$ does not wind ($n=0$).
Path $C$ can be shrunk to a point whereas the others cannot.  Each
path is said to belong to a different homotopy class.  Only paths $A$
and $B$ are affected by the topology of the space.  Path $C$ might
imagine that it is living in a space with no hole.\label{fig:2dhomo}}}

An example of the situation described above is the well-known $1+1$
dimensional sine-Gordon model defined by the Lagrangian ${\cal L}_{\rm
SG}=\frac{1}{2}(\partial_{\mu}\phi )^2 -\lambda\cos(\phi)$.  Here, the
topological current $q_\mu =
\frac{1}{2\pi}\epsilon_{\mu\nu}\partial^{\nu}\phi$ and is related to
the $\phi\rightarrow \phi+2\pi n$ symmetry of the ground state.  The
well known soliton and antisoliton (kink) solutions are absolutely
stable objects with topological charges $Q=\pm1$.  These cannot be
recovered by the standard methods of quantum field theory when one
starts from the vacuum state $\langle \phi\rangle =0$ and ignores the
topology.

In our case an analogous ground state symmetry is realized by the
procedure of summation over $l$ in Equation~(\ref{eq:Leff}), which
makes symmetry $\theta\rightarrow\theta+2\pi n$ explicit.  Thus, in
$1+1$ spatial dimensions we have the analogous stable objects.  The
fact that we actually consider $3+1$ dimensions means that the objects
are not point-like solitons as they are in the sine-Gordon model, but
rather, are two-dimensional domain walls with finite surface tension.

\subsection{Heavy degrees of freedom}
\label{sec:heavy}

What we have said up to this point is well known.  In the sine-Gordon
model, the solitons are absolutely stable objects as can be seen by
the fact that they are associated with the conserved current
$\frac{1}{2\pi}q_\mu = \epsilon_{\mu\nu}\partial^\nu \phi$ (here, the
indices run over the $1+1$ dimensions $z$ and $t$).  The conservation
is trivial: $\partial^\mu q_\mu = \frac{1}{2\pi}\epsilon_{\mu\nu}
\partial^\mu\partial^\nu \phi = 0$.  The corresponding topological
charge is $Q = \int q_t dz = \frac{1}{2\pi}\int -\partial_z\phi(z) dz
= \frac{1}{2\pi}(\phi_{z=-\infty}-\phi_{z=+\infty}) = n_--n_+$ which
is described by the winding number $n = n_+-n_-$.  Here, the field is
in a vacuum state $\phi_{z=\pm\infty} = 2\pi n_\pm$ at infinity.
Thus, we see that the charge is absolutely conserved and is integral.
  
In our effective theory (\ref{eq:Leff}), we consider an analogous
conserved current
\begin{equation}
\label{eq:current}
2\pi q_\mu = -i\epsilon_{\mu\nu} {\rm Tr}~ U^\dagger \partial^\nu U =
\epsilon_{\mu\nu} \partial^\nu \sum_i \phi_i = \epsilon_{\mu\nu}
\partial^\nu \phi_S\,,
\end{equation}
where we have introduced the notation $\phi_S = \sum_i \phi_i \sim
\eta'$ that we shall use later for the isotopical singlet ($\eta'$)
field.  By the same argument, we see that the two dimensional domain
walls of the theory~(\ref{eq:Leff}) are absolutely
stable.\footnote{One might think that, since the domain walls directly
involve the $\eta'$ field, that the stability of the $\eta'$ particle
might affect the stability of the domain walls.  This is not so.  Even
in the effective theory~(\ref{eq:Leff}), the $\eta'$ particle can be
considered as unstable decaying $\eta'\rightarrow 2\gamma$ for
instance.  This instability is related to the fact that the $\eta'$
number charge is not conserved. Irrespective of this non-conservation,
the current~(\ref{eq:current}) is still perfectly conserved.  The only
way for these domain walls to decay is by violating the conservation
of this current.  This is what we consider next.}

In the sine-Gordon model, this is the end of the story: the solitons
are absolutely stable.  In our effective Lagrangian~(\ref{eq:Leff}),
however, we have neglected the gluon degrees of freedom. In reality,
however, the gluon degrees of freedom are not very heavy.  Thus, we
must consider these extra degrees and look at how they affect the
charge conservation.  What we find is that, when we account for the
extra gluon degrees of freedom, the topology of the fields is no
longer restricted to the $\UU(1)$ manifold.  These extra degrees of
freedom allow the domain walls to continuously deform and to decay so
that the ground state exists everywhere.

To see how an extra degree of freedom can change the topology,
consider figure~\ref{fig:3dhomo}.  Here we have added a third
dimension to show that the barrier was actually a peg of finite
height.  Now that we can move in the extra dimension, we can use this
degree of freedom to ``lift'' the paths over the peg: thus, they are
no longer topologically stable.  In the QCD analogue, paths $A$ and
$B$ represents domain walls (path $C$ would is a trivial closed loop
which would relax to a point representing the same vacuum state
everywhere with no domain wall.).  There is an energy cost to ``lift''
the path over the barrier and at low temperatures $T\ll \Lambda_{{\rm
QCD}}$ there is not enough energy to do this, so classically, the
domain walls are stable.  It is still possible, however, for the walls
to overcome the barrier by tunnelling through the barrier.  The
tunnelling probability, however, could be low due to the height of the
obstacle and hence the lifetime of the walls could be much larger than
the $\Lambda_{{\rm QCD}}^{-1}$ scale which one might na\"\i{}vely
expected for standard QCD fluctuations.  See Section~\ref{sec:decay}
for details of the dynamics of the gluon~fields.

We should also remark that with each winding, the domain walls become
more energetic.  Walls with a large number of windings either have
enough energy to rapidly unwind, or else separate spatially forming
several domain walls of winding number $\sim \pm 1$.  For this reason,
we shall discuss in this paper only the simplest walls which wind
once.

\FIGURE[t]{\centerline{\epsfig{file=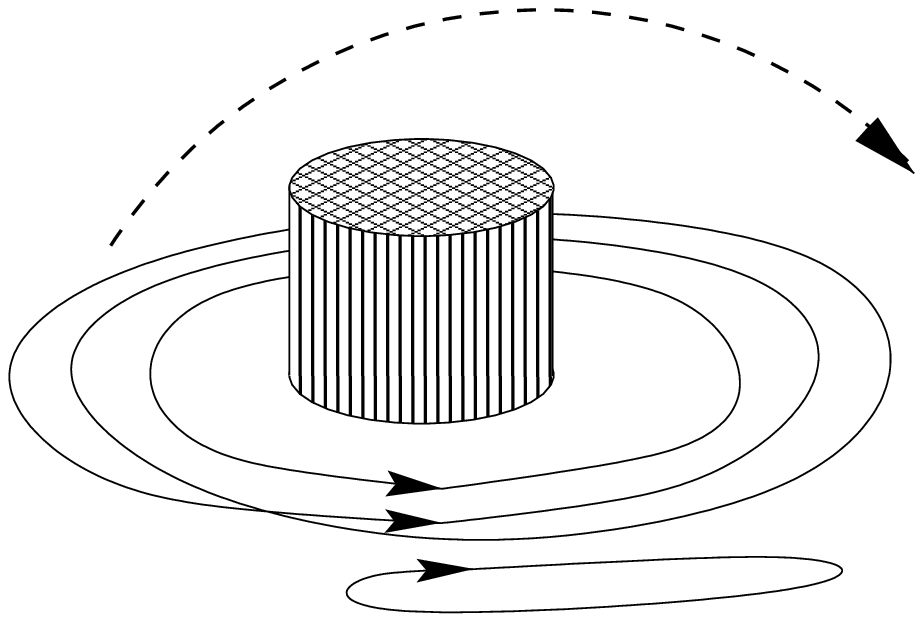,width=0.55\textwidth}}
\caption{Here we show the same picture as in figure~\ref{fig:2dhomo}
except that we show the third dimension.  Here we can see that all the
paths are now homotopically equivalent.  We can deform the paths by
``lifting'' them over the obstacle so that we can unwind them.  If the
paths were strings with some weight, then it would require some energy
to ``lift'' the strings over the obstacle.  If this energy was not
available, then we would say that, classically, the configurations
that wind around the peg are stable.  Quantum mechanically, however,
the strings could still tunnel through the peg, and so the
configurations are unstable quantum mechanically.  The probability
that one string could tunnel into another configuration would depend
on the height of the peg.~\label{fig:3dhomo}}}

To tie this picture together, consider the formerly conserved
current~(\ref{eq:current}) and effective Lagrangian~(\ref{eq:Leff})
and ask: from where do the phase fields $\phi_i$ come?  These phases
arise from some sort of complex field $\Phi = \rho e^{i\phi_S}$.  In
general, we must consider the dynamics, not only of the phase
$\phi_S$, but of the component $\rho = |\Phi|$.  We assume that this
field lives in some sort of Mexican-hat potential with approximate
symmetry $\phi_S\rightarrow \phi_S+\alpha$ and minimum valley where
$\langle \rho \rangle = 1$.  This symmetry is thus spontaneously
broken by the vacuum expectation values of the condensate
$\langle\Phi\rangle = \exp(i\langle\phi_S\rangle)$.  To recover the
effective Lagrangian: we ``integrate'' over the ``heavy'' $\rho$
degree of freedom by setting $\rho = \langle\rho\rangle = 1$ equal to
its classical expectation value and ignoring the quantum fluctuations
about this minimum.  This must reproduce the effective
potential~(\ref{eq:Leff}) with the pseudo-Goldstone field $\phi_S$
($\eta'$ and $\pi_i$ in the real theory).  The appropriate Mexican-hat
potential reproducing (\ref{eq:Leff}) is given in (\ref{d6}).

One can now consider an appropriate generalization of the
current~(\ref{eq:current})
\begin{equation}
2\pi q_\mu = -i \epsilon_{\mu\nu}{\rm Tr}~ \Phi^\dagger \partial^\nu
\Phi = \epsilon_{\mu\nu}\left(-i\rho \partial^\nu \rho +
\rho^2\partial^\nu \phi_S\right)
\end{equation}
which reduces to~(\ref{eq:current}) when we set $\rho =
\langle\rho\rangle = 1$.  Now we have
\begin{equation}
\pi \partial^\mu q_\mu = 
\rho \; \epsilon_{\mu\nu} \;\partial^\mu \rho \; \partial^\nu \phi_S
\end{equation}
which is no longer conserved as expected.  This was only conserved in
the effective theory~(\ref{eq:Leff}) because we integrated out the
heavy degrees of freedom by setting $\rho =1$ so that
$\partial^\mu\rho = 0$. Thus, the decay of the domain walls is
directly related to the dynamics of the heavy degrees of freedom.  We
consider this effect in Section~\ref{sec:decay}.  The physical object
responsible for this behaviour in~(\ref{eq:Leff}) is the gluon
condensate $E \sim \langle G^2\rangle$, which is the most essential
contribution to the mass $m_{\eta'}$ of the $\eta'$
particle.\footnote{If one assumes that all of the $\eta'$ physics
comes exclusively from the phase of the chiral condensate rather than
from gluon condensate, then one might argue (using the linear sigma
model) that the walls are classically unstable.  Were this the case,
then $m_{\eta'}\sim m_q \rightarrow 0$ in the chiral limit and the
$\UU(1)$ problem would remain
unresolved~\cite{Witten:1980sp,DiVecchia:1980ve}.  Thus, we see the
importance of the gluon condensate $E \sim \langle G^2\rangle \gg m_q$
which ensures the classical stability of the domain walls for $N_c
\geq 3$ as explained in Section~\ref{sec:decay}.}

\section{Domain walls}
\label{sec:dwall}

In the rest of this paper we limit ourselves with the simplest case
$N_f=2$ and neglect the difference between $f_{\pi}$ and $f_{\eta'}$
which numerically are very close to each other.  To describe the basic
structure of the QCD domain walls as well as that of axion domain
walls we replace the parameter $\theta$ in Equation~(\ref{eq:Leff}) by
a dynamical axion field $\theta\rightarrow N_f a=2a$ (this corresponds
to the so-called $N=2$ axion model).  We also introduce here the
following dimensionless phases, $\phi_S$ describing the isotopical
``singlet'' field, and $ \phi_T$ describing the isotopical ``triplet''
field. These fields correspond to the dynamical $\eta'$ (singlet) and
pion $\pi^0$ (triplet) fields defined in~(\ref{1}).  In principle,
there are other dynamical fields corresponding to the remaining
$\SU(N_f)$ generators (such as the charged pion fields $\pi^\pm$), but
one can show that these fields do not contribute to the domain wall
background but simply remain in their vacuum states.  These
fluctuations affect the overall energy density, but do not affect the
properties of the domain wall such as the surface tension and so we
neglect these in what follows.  (See for example~\cite{Huang:1985tt}
where the $N_f=2$ case is explicitly considered and $\pi^\pm=0$ along
the entire domain wall.)
\begin{equation}
\begin{array}{rclcrcl}
\phi_S &=& \phi_u+\phi_d\,, &\qquad& \eta' &=&
\displaystyle\frac{f_{\pi}}{2\sqrt{2}} \phi_S\,,
\\[9pt]
\phi_T &=& \phi_u-\phi_d\,, &\qquad& \pi^0 &=&
\displaystyle\frac{f_{\pi}}{2\sqrt{2}} \phi_T\,.
\end{array}
\end{equation}
In what follows we also need to know the masses of the relevant fields
in terms of parameters of the effective potential (\ref{eq:Leff}):
\begin{equation}
m_{\pi}^2 = \frac{4M}{f_\pi^2}\,, \qquad
m_{a}^2 = \frac{4M}{f_a^2}\left(1-\xi^2\right),
\end{equation}
and the $\eta'$ mass relation
\begin{equation}
\label{meta}
m_{\eta'}^2 = \frac{8}{N_c^2f_{\pi}^2}E+\frac{4M}{ f_{\pi}^2}\,,
\end{equation}
follows from~(\ref{13}) with $N_f=2$.  Here we have neglected all
possible mixing terms, and have introduced the following notations
\begin{eqnarray}
M &\equiv& \frac{M_u+M_d}{2} = \frac{ m_u\big|\!\langle 0|\bar{\Psi}_u \Psi_u|0 \rangle\!\big| 
+ m_d\big|\!\langle 0|\bar{\Psi}_d \Psi_d|0 \rangle\!\big| }{2}\,,
\nonumber \\
\xi &=& \frac{M_d-M_u}{M_d+M_u}\approx0.3\,.
\end{eqnarray}

\subsection{Domain wall equations}

To study the structure of the domain wall we look at a simplified
model where one half of the universe is in one ground state and the
other half is in another.  The fields will orient themselves in such a
way as to minimize the energy density in space, forming a domain wall
between the two regions.  In this model, the domain walls are planar
and we shall neglect the $x$ and $y$ dimensions for now.  Thus, a
complete description of the wall is given by specifying the boundary
conditions and by specifying how the fields vary along $z$.

The contribution of the light degrees of freedom to the energy density
of a domain wall is given by the following expression\footnote{There
may be additional contributions to the wall tension from the heavy
degrees of freedom integrated out to obtain~(\ref{eq:Leff}).  These
will be discussed in Section~\ref{sec:string-wall-tensions}} 
\wide{
  \begin{equation}
    \label{tension}
    \sigma = \int_{-\infty}^{\infty}\!\!\!\!\!\!{\rm d}z\left(
      \frac{f_{a}^2\dot{a}^2}{4}+ \frac{f_{\pi}^2\dot{\phi}_T^2}{16}+
      \frac{f_{\pi}^2\dot{\phi}_S^2}{16}+ V(\phi_S, \phi_T, a)-V_{\rm min}
    \right),
  \end{equation}
  }{
  \begin{equation}
    \label{tension}
    \sigma = \int_{-\infty}^{\infty}{\rm d}z\left(
      \frac{f_{a}^2\dot{a}^2}{4}+
      \frac{f_{\pi}^2\dot{\phi}_T^2}{16}+
      \frac{f_{\pi}^2\dot{\phi}_S^2}{16}+
      V(\phi_S, \phi_T, a)-V_{\rm min}
    \right)
  \end{equation}
  }
where the first three terms are the kinetic contribution to the energy
and the last term is the potential.  The kinetic term is actually a
four divergence, but we have assumed the wall to be a stationary
solution --- hence the time derivatives vanish --- and symmetric in
the $x$--$y$ plane.  The only dependence remaining is the $z$
dependence.  Here, a dot signifies differentiation with respect to
$z$: $\dot{a}=\frac{{\rm d}a}{{\rm d}z}$.

Now, to find the form of the domain walls, it is convenient to use a
form of the potential which follows from (\ref{7}):
\wide{
  \begin{equation}
    \label{eq:Vfull}
    V(\phi_S, \phi_T, a)=-E\cos\frac{\phi_S}{N_c}
    -2M\left(\cos\frac{\phi_T}{2} \cos\left(\frac{\phi_S}{2}+a\right)
      +\xi\sin\frac{\phi_T}{2} \sin\left(\frac{\phi_S}{2}+a\right) \right).
  \end{equation}
}{
\begin{equation}
  \label{eq:Vfull}
  V(\phi_S, \phi_T,a)
  =-2M\left(
    \cos\Bigl(\frac{\phi_S}{2}+a\Bigr)\cos\frac{\phi_T}{2} 
    +\xi\sin\Bigl(\frac{\phi_S}{2}+a\Bigr)\sin\frac{\phi_T}{2} 
  \right)-E\cos\frac{\phi_S}{N_c}.
\end{equation}
} Here we have redefined the fields $ \phi_u\rightarrow\phi_u+a$ and
$\phi_d\rightarrow\phi_d+a$ in order to remove the axion field from
the last term $\sim E$ and to insert it into the term $\sim M$.  To
minimize these equations, we can apply a standard variational
principle and arrive at the following equations of motion for the
domain wall solutions:
\wide{
  \begin{eqnarray}
    \label{eq:eqma}
    \frac{\ddot{a} f_a^2}{4M} &=&
    \cos\!\left(\frac{\phi_T}{2}\right)\sin\!\left(\frac{\phi_S}{2} +a
    \right)-\xi\sin\!\left(\frac{\phi_T}{2}\right)\cos\!\left(\frac{\phi_S}{2}+a
    \right),\\
    \label{eq:eqmpi}
    \frac{\ddot{\phi_T } f_\pi^2}{8M} &=&
    \sin\!\left(\frac{\phi_T}{2}\right)
    \cos\!\left(\frac{\phi_S}{2}+a\right)-\xi\cos\!\left(\frac{\phi_T}{2}\right)
    \sin\!\left(\frac{\phi_S}{2}+a\right) ,\\
    \label{eq:eqmeta'}
    \frac{\ddot{\phi_S} f_\pi^2}{8M} &=&
    \cos\!\left(\frac{\phi_T}{2}\right)
    \sin\!\left(\frac{\phi_S}{2}+a\right)-\xi\sin\!\left(\frac{\phi_T}{2}\right)
    \cos\!\left(\frac{\phi_S}{2}+a\right) +\frac{E}{M
      N_c}\sin\!\left(\frac{\phi_S}{N_c}\right),\qquad
  \end{eqnarray}
}{
\begin{align}
  \label{eq:eqma}
  \frac{\ddot{a} f_a^2}{4M} &=
  \cos\!\left(\frac{\phi_T}{2}\right)\sin\!\left(\frac{\phi_S}{2} +a \right)-\xi\sin\!\left(\frac{\phi_T}{2}\right)\cos\!\left(\frac{\phi_S}{2}+a \right),\\
  \label{eq:eqmpi}
  \frac{\ddot{\phi_T } f_\pi^2}{8M} &=
  \sin\!\left(\frac{\phi_T}{2}\right) \cos\!\left(\frac{\phi_S}{2}+a\right)-\xi\cos\!\left(\frac{\phi_T}{2}\right) \sin\!\left(\frac{\phi_S}{2}+a\right) ,\\
  \label{eq:eqmeta'}
  \frac{\ddot{\phi_S} f_\pi^2}{8M} &=
  \cos\!\left(\frac{\phi_T}{2}\right) \sin\!\left(\frac{\phi_S}{2}+a\right)-\xi\sin\!\left(\frac{\phi_T}{2}\right) \cos\!\left(\frac{\phi_S}{2}+a\right)
  +\frac{E}{M N_c}\sin\!\left(\frac{\phi_S}{N_c}\right),
\end{align}
}
where the last term of Equation~(\ref{eq:eqmeta'}) should be
understood as the lowest branch of the multivalued function described
by Equation~(\ref{eq:Leff}).  Namely, for $\phi_{S}\notin [0,\pi]$,
this should be interpreted as $\sin\bigl((\phi_S-2\pi l)/N_c\bigr)$
with the integer $l$ chosen to minimize the potential term
$\cos\bigl((\phi_S-2\pi l)/{N_c}\bigr)$.  For example, with $\pi \leq
\phi_S \leq 2\pi$, the last term should be of the form
$\sin\bigl((\phi_S -2\pi)/N_c\bigr)$.

Notice the following features: first, the trigonometric terms on the
right hand side are of, at most, order $1$; thus the scale for the
curvature (or rather, the second derivative) of the domain wall
solutions is limited by $f_a^2/M$ and $f_\pi^2/M$ etc.  In particular,
the axion domain wall must have a characteristic scale larger than
$m_a^{-1}/(1-\xi^2)$ and the pion domain wall must have a scale larger
than $m_\pi^{-1}$.  The last term in equation governing the $\phi_S$
field can potentially be somewhat larger than $1$, hence the smallest
scale for the $\phi_{S}$ field is related to the $\eta'$ mass.  We see
immediately that an axion domain wall must have a structure some
thirteen orders of magnitude larger than the natural QCD scale and
that the $\eta'$ field can have structure one order of magnitude
smaller than that of the pion field.

\subsection{QCD domain walls}
\label{sec:QCDwalls}

Here we consider the most important case of the QCD domain wall
solution which exists with or without an axion field.  So we now set
$a=0$.  The equations of motion become: 
\wide{
  \begin{eqnarray}
    \frac{\ddot{\phi_T } f_\pi^2}{8M} &=&
    \sin\left(\frac{\phi_T}{2}\right)
    \cos\left(\frac{\phi_S}{2}\right)-\xi\cos\left(\frac{\phi_T}{2}\right)
    \sin\left(\frac{\phi_S}{2}\right)\,,
    \nonumber\\
    \frac{\ddot{\phi_S} f_\pi^2}{8M} &=& \cos\left(\frac{\phi_T}{2}\right)
    \sin\left(\frac{\phi_S}{2}\right)-\xi\sin\left(\frac{\phi_T}{2}\right)
    \cos\left(\frac{\phi_S}{2}\right) +\frac{E}{M
      N_c}\sin\left(\frac{\phi_S}{N_c}\right)\,.\qquad
    \label{eq:eqmeta}
  \end{eqnarray}
}{
\begin{align}
  \label{eq:eqmeta}
  \frac{\ddot{\phi_T } f_\pi^2}{8M} &=
  \sin\!\left(\frac{\phi_T}{2}\right) \cos\!\left(\frac{\phi_S}{2}\right)-\xi\cos\!\left(\frac{\phi_T}{2}\right) \sin\!\left(\frac{\phi_S}{2}\right) ,
  \nonumber \\
  \frac{\ddot{\phi_S} f_\pi^2}{8M} &=
  \cos\!\left(\frac{\phi_T}{2}\right) \sin\!\left(\frac{\phi_S}{2}\right)-\xi\sin\!\left(\frac{\phi_T}{2}\right) \cos\!\left(\frac{\phi_S}{2}\right)
  +\frac{E}{M N_c}\sin\!\left(\frac{\phi_S}{N_c}\right).
\end{align}
} 
For convenience, we shall label the vacuum states using the notation
$(\phi_u,\phi_d)$.  Thus, we have only one physical ground state
$(\phi_u,\phi_d)=(0,0)$, however, because of the conserved topological
current~(\ref{eq:current}), classically stable domain walls can form
and interpolate from the ground state $(\phi_u,\phi_d)=(0,0)$ along a
path which is not homotopic to the null path.  To classify the paths
we use the redundant notation where $(0,0)$ and $(2\pi,0)$ etc. are
considered as different states and we talk about the field
interpolating between these states.  Keep in mind that this is only a
way of classifying the homotopy classes and that in fact all the
states represented by $(2\pi n,2\pi m)$ for integers $m$ and $n$ are
one in the same vacuum state.

The simplest domain wall is described by a continuous transition from
the ground state $(\phi_u,\phi_d)=(0,0)$ to the state labelled
$(\phi_u,\phi_d)=(2\pi ,0)$ as described by the vacuum solution
Equation~(\ref{10}) with $\theta=2\pi$ (or, equivalently, $l=-1$ in
Equation~(\ref{eq:Leff})). This wall corresponds to a single winding
around the $\UU(1)$ manifold.  It is also possible to wind in the
opposite sense.  To summarize, the two topologically stable domain
walls of minimal energy correspond to one winding in each direction
and are classified by the transitions from $(\phi_u,\phi_d)=(0,0)$ to:
\begin{description}
\item[Soliton]{$(\phi_u,\phi_d)=(2\pi,0)$.}
\item[Antisoliton]{$(\phi_u,\phi_d)=(-2\pi,0)$.}
\end{description}
\looseness=-1The solutions which wind from $(0,0)$ to $(0,\pm 2\pi)$ are not
topologically distinct from these due to the other pion fields and
have a higher energy.  Note, however, that in the chiral limit
$m_u=m_d$ and $\xi=0$, thus the transitions to $(\phi_u,\phi_d)=(0,\pm
2\pi)$ have the same energy and there is a degeneracy.  If $m_u>m_d$,
then these transition in $\phi_{d}$ are the minimal energy solutions
and the $\phi_{u}$ solutions above become unstable. In reality
$m_d>m_u$, and the transitions to $(\phi_u,\phi_d)=(\pm 2\pi, 0)$ are
the only stable transitions.

The general case of Equations~(\ref{eq:eqmeta}) cannot be solved
analytically and we present the numerical solution of
Equation~(\ref{eq:eqmeta}) in figure~\ref{fig:QCDwall}.  In order to
gain an intuitive understanding of this wall, we examine the solution
in the limit $m_{\pi}\ll m_{\eta'}$. In this case, the last term
of~(\ref{eq:eqmeta}) dominates unless $\phi_S$ is very close to the
vacuum states, $\phi_S\simeq 2\pi n$.  Thus, the central structure of
the $\phi_S$ field is governed by the differential
equation:\vspace{-1pt minus 10pt}
\begin{equation}
\label{eq:QCDeta}
\ddot{\phi}_S =\frac{8E}{N_c
f_{\pi}^2}\sin\left(\frac{\phi_S}{N_c}\right).
\end{equation}

Now, there is the issue of the cusp singularity when $\phi_S=\pi$
because we change from one branch of the potential to another as
expressed in Equation~(\ref{eq:Leff}). By definition, we keep the
lowest energy branch, such that the right hand side of
Equation~(\ref{eq:QCDeta}) is understood to be the function
$\sin(\phi_S/N_c)$ for $ 0 \leq \phi_S \leq \pi $ and
$\sin\bigl((\phi_S -2\pi)/N_c\bigr)$ for $\pi \leq \phi_S \leq 2\pi$.
However, we notice that the equations of motion are symmetric with
respect to the centre of the wall (which we take as $z=z_0$), hence
$\phi_S=\pi$ only at the centre of the wall and not before, so we can
simply look at half of the domain, $z\in(-\infty,0]$, with boundary
conditions $\phi_S=0$ at $z=-\infty $ and $\phi_S=\pi$ at $z=0$.  The
rest of the solution will be symmetric with $\phi_S=2\pi$ at
$z=+\infty$.  Equation~(\ref{eq:QCDeta}) with the boundary conditions
above has the solution (recall that $\phi_S\equiv \phi_u+\phi_d$)
\wide{
  \begin{equation}
    \label{eq:QCDetasol}
    \phi_S (z) = \begin{cases}
      4N_c \tan^{-1} \left[ \tan \displaystyle\frac{\pi}{4 N_c} 
        e^{\mu (z - z_0)} \right], & z\leq z_0  \\   
      2\pi  - 4 N_c\tan^{-1} \left[ \tan \displaystyle\frac{\pi}{4 N_c} 
        e^{- \mu (z - z_0)} \right], & z \geq z_0
    \end{cases}
  \end{equation}
}{
\begin{equation}
  \label{eq:QCDetasol}
  \phi_S (z) \equiv \phi_u+\phi_d = \begin{cases}
    4N_c \tan^{-1}\! \left[ \tan \frac{\pi}{4 N_c} 
     ~e^{\mu (z - z_0)}\right], & z\leq z_0  \\   
    2\pi  - 4 N_c\tan^{-1}\! \left[ \tan \frac{\pi}{4 N_c} 
      ~e^{-\mu (z - z_0)} \right], & z \geq z_0
  \end{cases}
\end{equation}
}
where $ z_0 $ is the position of the centre of the domain wall and 
\begin{equation}
  \label{eq:QCDwidth}
  \mu \equiv \frac{2  \sqrt{2E}}{ N_c f_{\pi} }\,, 
  \qquad \lim_{m_q\rightarrow 0}\mu = m_{\eta'}
\end{equation} 
is the inverse width of the wall, which is equal to the $m_{\eta'}$
mass in the chiral limit (see Equation~(\ref{meta})).  Thus, we see
that the dynamics of the central portion of QCD domain walls is
governed by the $\eta'$ field.  We shall also refer to the $\phi_S$
transition~(\ref{eq:QCDetasol}) which occurs in several places (see
for example Section~\ref{sec:etawall}) as the $\eta'$ domain wall.
The first derivative of the solution is continuous at $ z = z_0 $, but
the second derivative exhibits a finite jump.

Before we continue our discussions regarding the structure of QCD
domain walls, a short remark is in order: The sine-Gordon equation,
which is similar to Equation~(\ref{eq:QCDeta}) with a cusp
singularity, was first considered in~\cite{Fugleberg:1998kk} where a
solution similar to~(\ref{eq:QCDetasol}) was presented. There is a
fundamental difference, however, between domain walls discussed
in~\cite{Fugleberg:1998kk} and the domain walls we consider here.
In~\cite{Fugleberg:1998kk}, the domain walls were constructed as
auxiliary objects in order to describe a decay of metastable vacuum
states which may exist under the certain circumstances.  The walls we
discuss here classically \emph{stable} physical objects where the
solutions interpolate between the same vacuum state; their existence
is a consequence of the topology of the $\UU(1)$ singlet $\eta'$
field.  This topology, represented by the exact symmetry $\theta\equiv
\theta+2\pi n$ in Equation~(\ref{eq:Leff}) is a very general property
of QCD and does not depend on the specific choice of parameters or
functional form of the effective potential.  Similar equations and
solutions with application to the axion physics were also discussed
in~\cite{Gabadadze:2000vw}.

The solution described above dominates on scales where $|z|\leq
\mu^{-1}$, however, the isotopical triplet pion transition can only
have a structure on scales larger than $m_\pi^{-1}\gg \mu^{-1}$ and so
the central structure of the $\eta'$ wall can have little effect on
the pion field.  Indeed, we can see that, for $|z|\gg \mu^{-1}$,
$\phi_S$ is approximately constant with the vacuum values.  Making
this approximation, we see that the isotopical triplet field is
governed by the equation
\begin{equation}
\label{eq:QCDpi}
\ddot{\phi}_T=2m_{\pi}^2\sin\!\left(\frac{\phi_T}{2}\right).
\end{equation}
This has the same form as~(\ref{eq:QCDeta}) and hence the solution is
(recall that $\phi_T \equiv \phi_u-\phi_d$) 
\wide{
  \begin{equation}
    \label{eq:QCDpisol}
    \phi_T (z) \simeq \begin{cases}
      8 \tan^{-1} \left[ \tan \displaystyle\frac{\pi}{8} 
        e^{m_{\pi} (z - z_0)}\right], & z \ll z_0 -\mu^{-1}
      \\[7pt]
      2\pi  - 8\tan^{-1} \left[\tan\displaystyle\frac{\pi}{8} 
        e^{- m_{\pi} (z - z_0)}\right], &
      z \gg z_0 + \mu^{-1} 
    \end{cases}
  \end{equation}
}{
\begin{equation}
  \label{eq:QCDpisol}
  \phi_T (z) \equiv \phi_u-\phi_d =
  \begin{cases}
    8 \tan^{-1}\!\left[ \tan\tfrac{\pi}{8} 
      ~ e^{m_{\pi} (z - z_0)} \right], &   z \ll z_0 -\mu^{-1}, \\   
    2\pi  - 8\tan^{-1}\!\left[ \tan\tfrac{\pi}{8} 
      ~ e^{- m_{\pi} (z - z_0)} \right], & z \gg z_0 + \mu^{-1},
  \end{cases}
\end{equation}
}
which is a reasonable approximation for all $z$. Numerical solutions
for the $\phi_S$ and $\phi_T$ fields are shown along with the same
solution in terms of the $\phi_u$ and $\phi_d$ in
figure~\ref{fig:QCDwall}.  As we can see from the explicit form of the
presented solution, the $\eta'$ transition is sandwiched in the pion
transition.  This is a key feature for some applications of this type
of the domain wall as discussed in~\cite{Forbes:2000gr} for example.

The contribution to the wall surface tension defined by
Equation~(\ref{tension}) can be easily calculated analytically in the
chiral limit when the analytical solution is known and is given by
Equations~(\ref{eq:QCDetasol}) and~(\ref{eq:QCDpisol}).  Simple
calculations leads to the following contribution from the pion and
$\eta'$ fields.
\begin{equation}
\label{a3}
\sigma=\frac{4 N_c}{\sqrt{2}} f_{\pi}\sqrt{\left\langle \frac{ b
\alpha_s }{ 32 \pi} G^2 \right\rangle } \, \left( 1 - \cos \frac{\pi}{
2 N_c} \right) + {\rm O}(m_q f_{\pi}^2 )\,.
\end{equation}
In case when $m_q\neq 0$, an analytical solution is not known, but
numerically, $\sigma$ is close to the estimate~(\ref{a3}).

\FIGURE[t]{\begin{centering}
\psfrag{eta}{}
\psfrag{cz}{\scriptsize$zm_{\eta'}$}
\psfrag{Pionxxx}{\scriptsize$\phi_{T}(z)$}
\psfrag{Etaxxx}{\scriptsize$\phi_{S}(z)$}
\psfrag{Phiuxxx}{\scriptsize$\phi_{u}(z)$}
\psfrag{Phidxxx}{\scriptsize$\phi_{d}(z)$}
\includegraphics[width=0.48\textwidth]{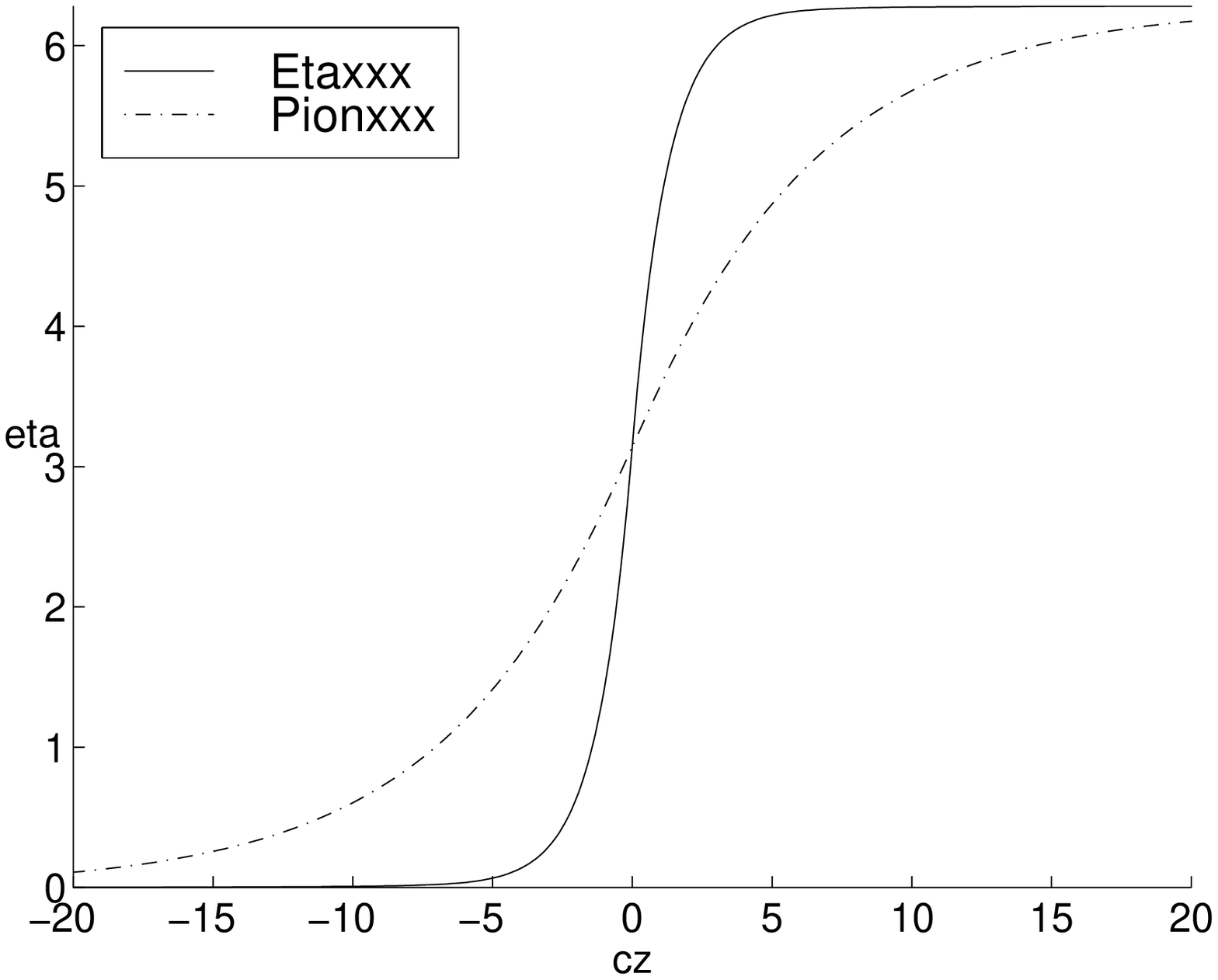}
\includegraphics[width=0.48\textwidth]{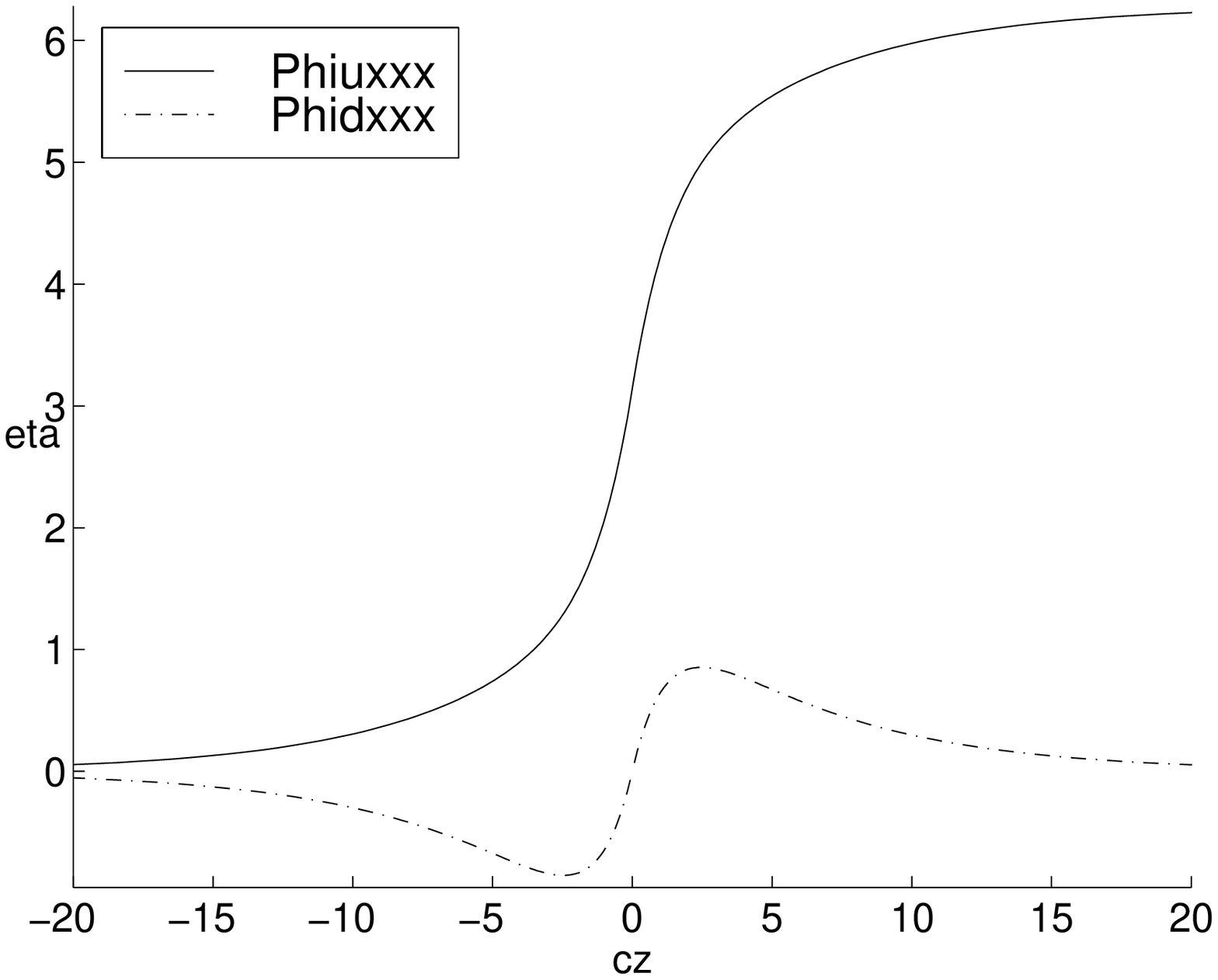}
\caption{Basic form of the QCD domain walls.  Notice that the
scale for the pion transition is larger than for the $\eta'$
transition and that the width of the $\eta'$ wall is set by the
scale $m_{\eta'}$.
\label{fig:QCDwall}}
\end{centering}}

\subsection{Axion dominated domain walls}
\label{sec:axionwalls}

In the previous subsection when the QCD domain walls were discussed,
the axion was not introduced as a dynamical field.  In this subsection
we assume that the axions exist.  In this case there are domain walls
in which the axion is the dominant player.  The introduction of
axions, in most cases, makes the domain wall an absolutely stable
object.  Our case is no exception and the axion model under
discussion, (which is the $N=2$ axion models according to the
classification~\cite{Kim:1987ax,Cheng:1988gp,Peccei:1989}) is an
absolutely stable object.  At the same time it is
well-known~\cite{Zeldovich:1974uw,Vilenkin:1994}, that stable domain
walls can be a cosmological disaster.  We do not address in this paper
the problem of avoiding a domain wall dominated universe.  Rather, we
would like to describe some new elements in the structure of axion
domain walls, which were not previously discussed.

The first and most natural type of the axion domain wall was discussed
by Huang and Sikivie~\cite{Huang:1985tt} who neglected the $\eta'$
field in their construction.  We shall refer to this wall as the
Axion-Pion domain wall ($a_{\pi}$).  As shown in~\cite{Huang:1985tt},
it has a width of the scale $\sim m_a^{-1}\gg \Lambda_{{\rm
QCD}}^{-1}$ for both the axion and $\pi$ meson components.  As Huang
and Sikivie expected, the $\eta'$ plays a very small role in this
wall.  In what follows we include a discussion of this type of domain
wall for the completeness.

\FIGURE{\begin{centering}
\psfrag{eta}{$$} \psfrag{caz}{\scriptsize$zm_{a}$} \psfrag{Axion}{\scriptsize$a(z)$}
\psfrag{Pion}{\scriptsize$\phi_{T}(z)$} \psfrag{Eta}{\scriptsize$\phi_{S}(z)$}
\includegraphics[width=0.55\textwidth]{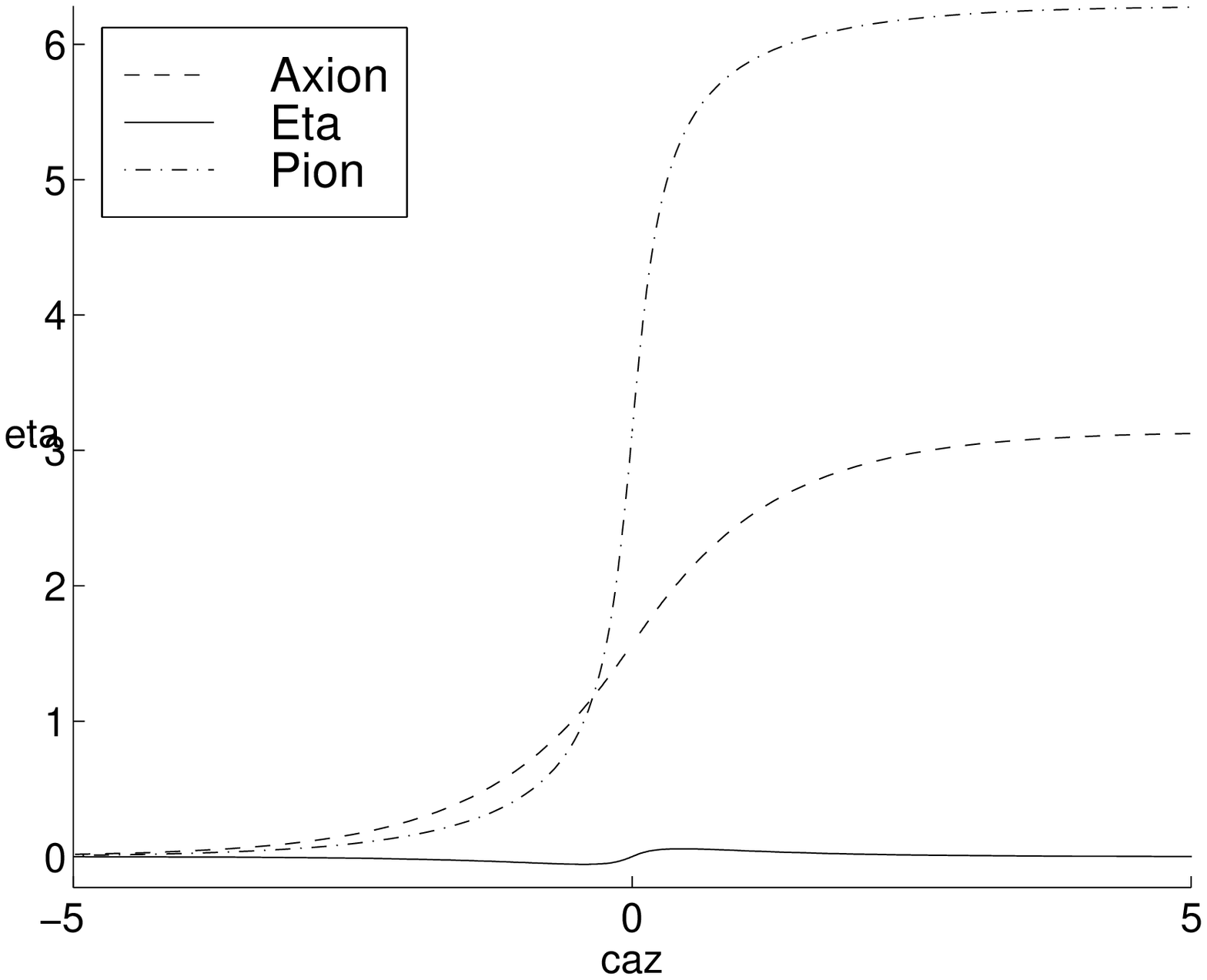}
\caption{Basic form of the $a_{\pi}$ domain wall.  Notice that
all the fields have a structure on the scale of $m_a^{-1}$ and
that the isotopical singlet fields plays a very small role.
\label{fig:apwall1}}
\end{centering}}

Our original result is to description a new type of the axion domain
wall in which the $\eta'$ field is a dominant player.  We shall call
this new solution the Axion-Eta' domain wall ($a_{\eta'}$). This new
type of the domain wall was considered for the first time
in~\cite{Forbes:2000gr} as a possible source for galactic magnetic
fields in early universe.  In what follows we give a detail
description of the solution for the $a_{\eta'}$ wall.  Here we want to
mention the fundamental difference between the $a_{\pi}$ wall
discussed in~\cite{Huang:1985tt} and the $a_{\eta'}$ wall introduced
in~\cite{Forbes:2000gr}. 

{\sloppy
Unlike the $a_{\pi}$ wall which has structure only on the huge scale
of $m_a^{-1}$, the $a_{\eta'}$ wall has nontrivial structure at both
the axion scale $m_a^{-1}$ as well as at the QCD scale
$m_{\eta'}^{-1}\sim\Lambda_{\rm QCD}^{-1}$.  The reason for this is
that, in the presence of the non-zero axion field (which is equivalent
to a non-zero $\theta$ parameter), the pion mass is efficiently
suppressed due to its Goldstone nature, thus the pion field follows
the axion field and has a structure on the same $m_{a}^{-1}$
scale. The $\eta'$, however, is not very sensitive to $\theta$ and so
it remains massive.  

}
Again, the $a_{\eta'}$ solution has a sandwich structure with the
singlet transition occurring at the centre of the wall.  One can adopt
the viewpoint that the $a_{\eta'}$ domain wall is an axion domain wall
with a QCD domain wall sandwiched in the centre.  This phenomenon is
critical for applications involving the interaction of domain walls
with strongly interacting particles.  Indeed, there is no way for the
$a_{\pi}$ wall to trap any strongly interacting particles, like
nucleons, because of the huge difference in scales; the $a_{\eta'}$
wall, however, has a QCD structure and can therefore efficiently
interact with nucleons on the QCD scale $\Lambda_{\rm QCD}^{-1}$.
Regarding the wall tension~(\ref{tension}), it is dominated by the
axion physics and thus has the same order of magnitude for both types
of the axion domain walls which is proportional to $\sigma \sim M/m_a
\sim f_a f_{\pi}m_{\pi}$ as found in~\cite{Huang:1985tt}.

\subsubsection{Axion-pion domain wall}

The solution discussed by Huang and Sikivie corresponds to the
transition $(a,\phi_u,\phi_d):\;(0,0,0)\rightarrow(\pi,\pi,-\pi)$,
i.e., by a transition in the axion and pion fields only.  This
transition describes the $a_{\pi}$ wall.  Indeed, in terms of
$(a,\phi_S,\phi_T)$ this transition corresponds to a nontrivial
behaviour of the axion and triplet pion component $\phi_T$ of the
$\phi_u,\phi_d$ fields:
$(a,\phi_S,\phi_T):\;(0,0,0)\rightarrow(\pi,0,2\pi) $.  The other
transition (the $a_{\eta'}$ wall) corresponds to the transition
$(a,\phi_u,\phi_d):\;(0,0,0)\rightarrow(-\pi,\pi,\pi)$ in which the
singlet $\eta'$ field is dominant:
$(a,\phi_S,\phi_T):\;(0,0,0)\rightarrow(-\pi,2\pi,0) $.  It will be
considered later on.

Huang and Sikivie discussed the solution to this wall in the limit
where the $\eta'$ field is extremely massive and hence they neglected
its role. It can be integrated out which effectively corresponds to
fixing it $\eta'(z)=0$. Indeed, if we simply fix $\eta'(z)=0$ in our
equations, then we reproduce their solution.  When $E$ is large, then
$m_\pi\ll m_{\eta'}$ as Huang and Sikivie assumed, the effects of the
$\eta'$ particle can be neglected and the solution for the the axion
and pion fields presented in~\cite{Huang:1985tt} is valid for the
boundary conditions described above.  We plot this numerical solution
which includes the $\eta'$ effects in figure~\ref{fig:apwall1}. As
announced above, the $a_{\pi}$ solution has the only scale of $ \sim
m_a^{-1}$ for both components, axion as well as the pion field $ \sim
\phi_T$.  The $\eta'$ field remains close to the its vacuum value and
only slightly corrects the solution.

\subsubsection{Axion-eta' domain wall}
\label{sec:etawall}

Having looked at the solution of the $a_{\pi}$ wall, we now
investigate the structure of the $a_{\eta'}$ domain wall which is a
new solution.  This solution corresponds to the transition
$(a,\phi_u,\phi_d):\;(0,0,0)\rightarrow(\pi,\pi,\pi)$.  Now the
singlet $\eta'$ field undergoes a transition instead of the triplet
pion field: $(a,\phi_S,\phi_T):\;(0,0,0)\rightarrow(-\pi,2\pi,0)$.  As
we discussed above, the singlet field never becomes massless, and
therefore, a new structure at the QCD scale $\sim\Lambda_{{\rm
QCD}}^{-1}$ emerges in sharp contrast to the well-studied $a_{\pi}$
domain wall~\cite{Huang:1985tt} where no such structure appears.  We
should note that the potential (\ref{eq:Vfull}) has the same vacuum
energy at $\phi_S=0$ as well as at $\phi_S=2\pi$ due to the change of
$\phi_S$ field to the lowest energy branch at $\phi_S=\pi$ as
described by Equation~(\ref{eq:Leff}).  Therefore, this domain wall
interpolates between two degenerate states, and thus, like the
$a_{\pi}$ domain wall~\cite{Huang:1985tt}, the solution under these
considerations is absolutely stable.

For $|z|\gg \mu^{-1}$, the last term in~(\ref{eq:eqmeta'}) is
negligible and so the solution behaves like the $a_{\pi}$ wall.  What
happens is that, away from the wall, the axion field dominates and
shapes the wall as it does in with the $a_{\pi}$ solution.  Again, the
pion mass is suppressed and $\phi_S\approx 0$.  As $z\sim \mu^{-1}$
however, the last term of~(\ref{eq:eqmeta}) starts to dominate the
behaviour.  At this point, the $a_{\eta'}$ wall undergoes a sharp
transition similar to the QCD domain wall described by
(\ref{eq:QCDetasol}).  We plot this solution along with a blowup in
figure \ref{fig:aewall}.  Notice also, that the singlet field
cancels the effects of the axion near the center of the wall, and so
the pion field becomes massive again as it undergoes its transition.
\FIGURE[t]{\begin{centering}
\psfrag{eta}{$$}
\psfrag{cza}{\scriptsize$zm_{a}$}
\psfrag{cze}{\scriptsize$zm_{\eta'}$}
\psfrag{Axion}{\scriptsize$a(z)$}
\psfrag{Pion}{\scriptsize$\phi_{T}(z)$}
\psfrag{Eta}{\scriptsize$\phi_{S}(z)$}
\includegraphics[width=0.48\textwidth]{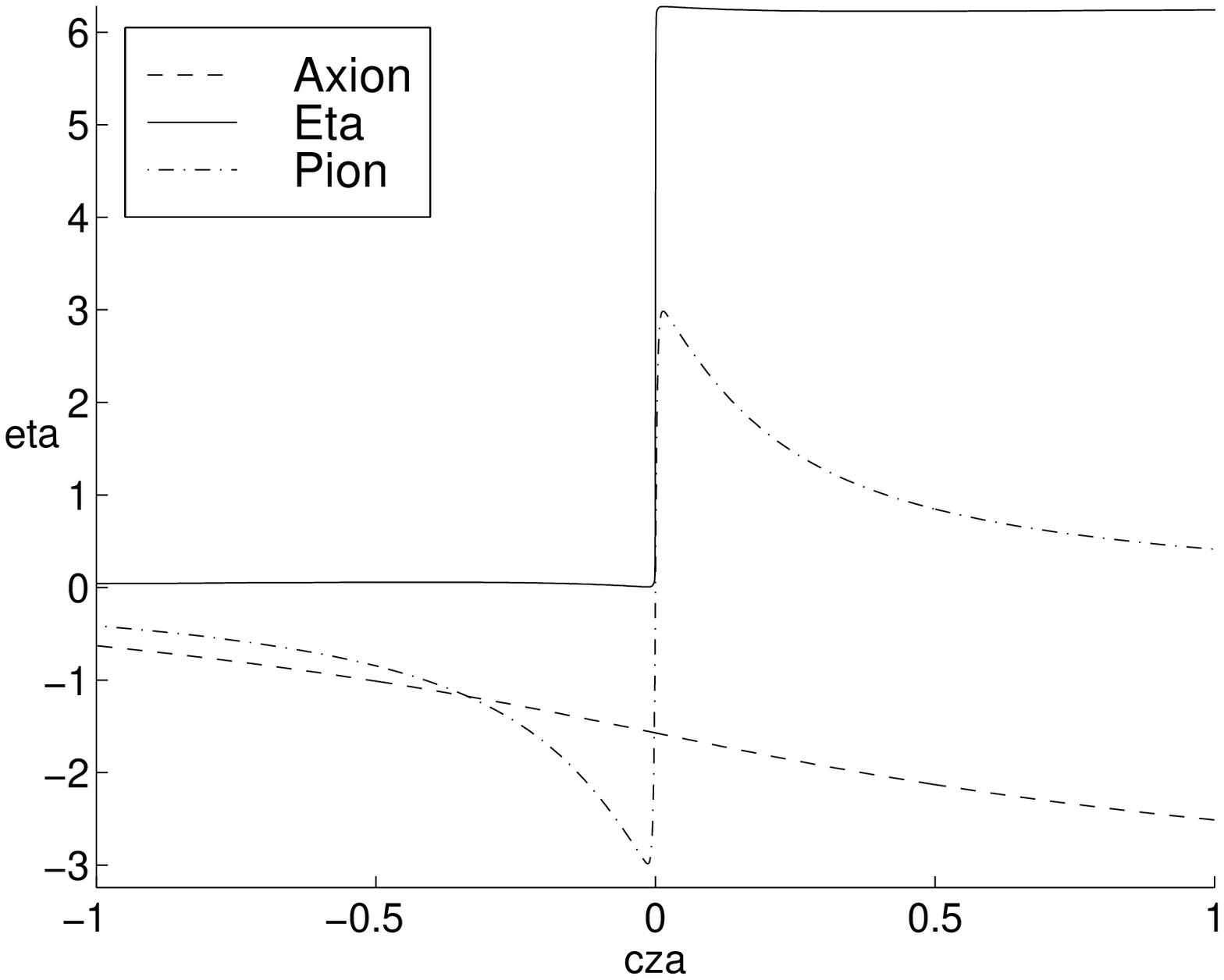}
\includegraphics[width=0.48\textwidth]{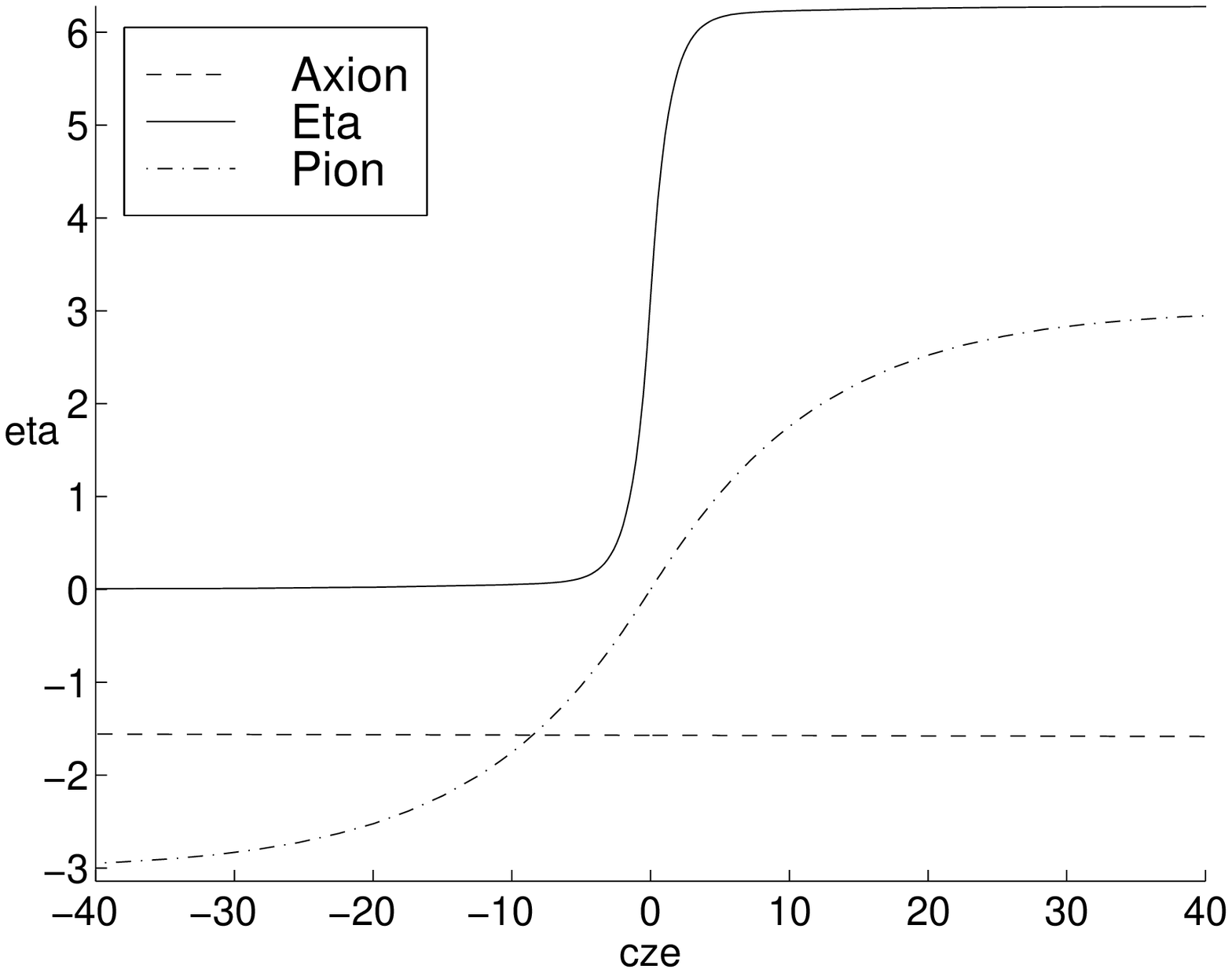}
\caption{Basic form of the $a_{\eta'}$ domain walls with a closeup
where the axion field $a\approx\pi/2$.  Notice that the large scale
structure is similar to that of the $a_{\pi}$ wall, but that there is
also a small scale structure on the scale of $m_{\eta'}$.  Near the
centre of the wall, the pion regains its mass and undergoes a
transition on the scale of $m_{\pi}$.  \label{fig:aewall}}
\end{centering}}

\section{Decay of the QCD domain walls}
\label{sec:decay}

We do not have much new to say regarding the generalities of domain
walls, nor do we have a resolution for the general problem of avoiding
a domain wall dominated universe.  We have nothing new to say
regarding the stability or evolution of axion domain
walls~\cite{Sikivie:1982qv}--\cite{Hagmann:1991mj}
(which were also discussed in the previous sections),
see~\cite{Chang:1998tb} and references therein for a recent review on
the subject.  We refer the reader to the nice text
book~\cite{Vilenkin:1994} for general discussion about domain walls,
other topological defects and their role in the early universe.  This
section is devoted specifically to the QCD domain walls discussed in
Section~\ref{sec:QCDwalls}.

Up to now, we have treated the domain walls as topologically stable
objects.  If we only consider the low energy degrees of freedom
present in the effective theory~(\ref{eq:Leff}) then this is certainly
the case as discussed in Section~\ref{sec:topology}.  We shall also
argue that this picture of classically stable domain walls is correct
in the appropriate large $N_c=\infty$ and chiral $m_q=0$ limit.  In
reality, however, one must consider the heavy degrees of freedom that
were integrated out to obtain~(\ref{eq:Leff}). For finite $N_c$, one
finds that QCD domain walls (not axion domain walls) are unstable on
the quantum level due to a tunnelling mode of decay as described in
Section~\ref{sec:heavy}. We estimate the relevant life-time of these
walls with respect to this quantum transition and show that, although
the walls don't live long on a cosmological scale, exponential
suppression of the tunnelling decay mode may result in the lifetime of
the walls being much larger than the QCD scale $\Lambda_{{\rm
QCD}}^{-1}$.  We argue that this exponential suppression remains in
the large $N_c$ limit where we have theoretical control, and may also
occur numerically for physical values of the parameters.  Therefore,
these walls do not pose a cosmological problem as one might
na\"\i{}vely suspect.\footnote{We should remind the reader once again
that the existence of the QCD domain walls described above is a
consequence of the well-understood symmetry
$\theta\rightarrow\theta+2\pi n$ and is a consequence of the
topological charge quantization; their existence is not based on any
model-dependent assumptions we have made to support the specific
calculations in the previous sections.  The question now is not the
existence of these walls, but whether or not they live long enough to
affect relevant physics.}

Even though QCD domain walls are ultimately unstable, they may still
play an important role in physical processes with timescales
comparable to the lifetime of the walls.  QCD domain walls are not
likely to exist today, however, they may play an important role
shortly after the QCD phase transition in the evolution of the early
universe, or in heavy ion collisions after the transition from a
quark-gluon plasma to the hadronic phase when system cools.  In these
cases, QCD domain walls with a finite lifetime may play an important
role.  Indeed, according to the standard theory of cosmological phase
transitions~\cite{Kibble:1976sj,Vilenkin:1994}, if, below a critical
temperature $T_c$, the potential develops a number of degenerate
minima, then the choice of minima will depend on random fluctuations
in fields.  The minima that the fields settle to can be expected to
differ in various regions space. If neighbouring volumes fall into
different minima, then a kink (domain wall) will form as a boundary
between them.\footnote{\label{foot:PT} Whether or not domain wall
actually will form at the QCD transition in the early universe is an
unresolved issue.  Presently it is believed that the QCD phase
transition is actually a smooth crossover (see~\cite{Rajagopal:2000wf}
for a review of the QCD phase diagram).  If this is the case, then it
is possible that no domain walls form because the universe cools very
slowly.  To resolve this question, one must estimate the relaxation
timescales involved: if the crossover is sharp enough, then domain
walls may still form.  At the RHIC, however, the system is quenched
due to the rapid expansion of the colliding ions and so the formation
of domain walls is very likely if an appropriate part of the phase
diagram is explored by the reaction.}  The relevant question becomes:
``Is the lifetime of QCD domain walls large enough to be of physical
interest?''  In what follows, we hope to convince the reader that the
answer to this question may in fact be: ``Yes!''.

\subsection{Estimating the decay rate}
\label{sec:estimates}

In the following, we estimate the lifetime of the QCD domain walls due
to a tunnelling process whereby a hole forms in the domain wall and
expands, consuming the wall.  A quantitative calculation of this rate
is presently beyond our control, but we can estimate the magnitudes of
the effect through a semi-classical approximation.  In
Section~\ref{sec:large-n_c-limit} we shall argue that these
approximations are valid asymptotically in the large $N_c$ limit and
that the decay rate is exponentially suppressed.

The decay mechanism is due to a tunnelling process which creates a
hole in the domain wall which connects the $(\phi_{u},\phi_{d})=(0,0)$
domain on one side of the wall to the $(\phi_{u},\phi_{d})= (2\pi,0)$
domain on the other (see Equation~(\ref{10})).  Passing through the
hole, the fields remain in the ground state.  This lowers the energy
of the configuration over that where the hole was filled by the domain
wall transition by an amount proportional to $R^2$ where $R$ is the
radius of the hole.  The hole, however, must be surrounded by a
string-like field configuration.  This string represents an excitation
in the heavy degrees of freedom and thus costs energy, however, this
energy scales linearly as $R$.  Thus, if a large enough hole can form,
then it will be stable and the hole will expand and consume the wall.

This process is commonly called quantum nucleation and is similar to
the decay of a metastable wall bounded by strings, and we use a
similar technique to estimate the tunnelling probability.  The idea of
the calculation was suggested by Kibble~\cite{Kibble:1982dd}, and has
been used many times since then (see the textbook~\cite{Vilenkin:1994}
for a review).  The most well known example of such a calculation is
the calculations of the decay rate in the so-called $N_{PQ}=1$ axion
model where the axion domain wall become unstable for a similar reason
due to the presence of axion
strings~\cite{Davis:1985pt}--\cite{Chang:1998tb}.  However, as was
emphasized in~\cite{Dvali:1995wv}, the existence of strings as the
solutions to the classical equations of motion is not essential for
this decay mechanism (see below).  Some configurations, not
necessarily the solutions of classical equations of motion, which
satisfy appropriate boundary conditions, may play the role played by
strings in the $N_{PQ}=1$ axion model.

To be more specific, let us consider a closed path starting in the
first domain with $(\phi_u,\phi_d)=(0,0)$, which goes through the hole
and finally returns back to the starting point by crossing the wall
somewhere far away from the hole.  The phase change along the path is
clearly equals to $ \phi_u+\phi_d=2\pi$.  Therefore, the absolute
value of a field which gives the mass to the $\eta'$ field (the
dominant part of the domain wall) has to vanish at some point inside
the region encircled by the path. By moving the path around the hole
continuously, one can convince oneself that there is a loop of a
string-like configuration (where the absolute value of a relevant
field vanishes such that the $\eta'$ singlet phase is a well defined)
enclosing the hole somewhere. In this consideration we did not assume
that a hole, or string enclosing the hole, are solutions of the
equations of motion.\footnote{\label{fn7}It is quite obvious that such
a configuration cannot be described within our non-linear $\sigma$
model given by Equation~(\ref{eq:Leff}) where it was assumed that the
gluon as well as the chiral condensates are non-zero constants.  In
this case, the singlet phase is not well defined everywhere.  However,
in the case of a triplet $\pi$ meson string, such a configuration can
easily be constructed within a linear $\sigma$ model by allowing the
absolute value of the chiral condensate to fluctuate along with the
Goldstone phase ($\pi$ meson field). The $\sigma$ term in the linear
$\sigma$ model essentially describes the rigidity of the potential.
Indeed, the corresponding calculations within a linear $\sigma$ model
were carried out in~\cite{Brandenberger:1998ew} where it was
demonstrated that the solution describing the $\pi$ meson string
exists, albeit unstable as expected from the topological arguments.
To carry out a similar calculations in our case for the singlet
$\eta'$ phase, one should allow fluctuations of the gluon fields: the
fields that give mass to $\eta'$ meson and that describe the rigidity
of the relevant potential. (See~(\ref{eq:current}) and the surrounding
discussion.)}  They do not have to be solutions.

However, if we want to describe the hole nucleation
semi-classically~\cite{Kibble:1982dd,Vilenkin:1994}, then we should
look for a corresponding instanton which is a solution of Euclidean
(imaginary time, $t=i\tau$) field equations, approaching the
unperturbed wall solution at $\tau\rightarrow \pm\infty$.  In this
case the probability $P$ of creating a hole with radius $R$ per area
$S$ per time $T$ can be estimated as follows\footnote{The estimate
given below is designed for illustrative purposes only, and should be
considered as a very rough estimation of the effect to an accuracy not
better than the order of
magnitude.}~\cite{Kibble:1982dd,Kobzarev:1975cp,Coleman:1977py}:
\begin{equation}
\label{d3}
\frac{P}{ST}\sim\left[\sqrt{\frac{S_0}{2\pi}}\right]^{3}e^{-S_0}
\times {\rm Det}\,,
\end{equation}
where $S_0$ is the classical instanton action; Det can be calculated
by analyzing small perturbations (non-zero modes contribution) about
the instanton, (see~\cite{Coleman:1977py} for an explanation of the
meaning of this term) and will be estimated using dimensional
arguments; and $\bigl(\sqrt{S_0/(2\pi)}\bigr)^{3}$ is the contribution
due to three zero modes describing the instanton
position.\footnote{The three zero modes in our case should be compared
with the four zero modes from the calculations
of~\cite{Coleman:1977py}.  This difference is due to the fact that
in~\cite{Coleman:1977py} the decay of three dimensional metastable
vacuum state was discussed.  In our case, we discuss a decay of a
two-dimensional object.}

If the radius of the nucleating hole is much greater than the wall
thickness, we can use the thin-string and thin-wall approximation.
(The critical radius $R_c$ will be estimated later and this
approximation justified).  In this case, the action for the string and
for the wall are proportional to the corresponding worldsheet areas
\cite{Kibble:1982dd},
\begin{equation}
\label{d4}
S_0=4\pi R^2\alpha -\frac{4}{3}\pi R^3\sigma\,.
\end{equation}
The first term is the energy cost of forming a string: $\alpha$ is the
string tension and $4\pi R^2$ is its worldsheet area.  The second term
is energy gain by the hole over the domain wall: $\sigma$ is the wall
tension and $4/3\pi R^3$ is its worldsheet volume.  The world sheet of
a static wall lying in the $x$-$y$ plane is the three-dimensional
hyperplane $z=0$.  In the instanton solution, this hyperplane has a
``hole'' which is bounded by the closed worldsheet of the string.

Minimizing~(\ref{d4}) with respect to $R$ we find the critical radius
\begin{equation}
\label{d5}
R_c=\frac{2\alpha}{\sigma}\,,\qquad
S_0=\frac{16\pi\alpha^3}{3\sigma^2}\,.
\end{equation} 
The lorentzian evolution of the hole after nucleation can be found by
making the inverse replacement $\tau\rightarrow -it$ from Euclidean to
Minkowski space-time. The hole expands with time as $x^2+y^2=R^2+t^2$,
rapidly approaching the speed of light.

To estimate the appropriate string and wall tensions, we must step
back from the effective theory~(\ref{eq:Leff}) and include the heavy
degrees of freedom that allow the fields to tunnel.  Unfortunately, to
provide a well justified estimate of these parameters in standard QCD
is very difficult because we do not know how to quantitatively include
the effects of the heavy degrees of freedom.  We shall argue, however,
that we regain this theoretical control in the chiral and large $N_c$
limits.  We postpone this discussion until
Section~\ref{sec:heavy-degr-freed}, but we summarize the $N_c$
dependence of these parameters here:
\begin{equation}
\label{eq:tensions_Nc}
N_c \lesssim \alpha\,,\qquad   \sqrt{N_c} \lesssim \:\sigma\,.
\end{equation}
These lower bounds for the string tension $\alpha$ and wall tension
$\sigma$ are well justified.  For our argument, however, we need an
upper bound on the wall tension.  As we shall argue later, we expect
that the lower bound for $\sigma$ in~(\ref{eq:tensions_Nc}) is
actually the upper bound and that:
\begin{equation}
\label{eq:5}
\sigma\sim\sqrt{N_c}\,.
\end{equation}
However, there is some speculation about the correct upper bound and
one might argue that that the upper bound could be of order $N_c$.
While we strongly suspect that~(\ref{eq:5}) is correct, we cannot
prove this and thus consider the possibility of larger~$\sigma$.

In either case, from~(\ref{d5}) we see that, in the large $N_c$ limit,
the probability of producing a hole~(\ref{d3}) is exponentially
suppressed by the factor of at least $e^{-S_0} \sim e^{-N_c}$ in the
case of $\sigma\sim N_c$ (In this case, our semiclassical estimate of
$S_0$ is not numerically justified, however, we believe that it is
parametrically still valid as we shall discuss in
Section~\ref{sec:large-n_c-limit}).  We believe, however,
that~(\ref{eq:5}) holds and thus that the actual suppression is much
stronger
\begin{equation}
\label{eq:decay_rate_Nc}
e^{-S_0} \sim e^{-N_c^2}\,.
\end{equation}
The only way to kill the exponential suppression is to arrange for
$\sigma > N_c^{3/2}$ which, as we shall discuss, has no
phenomenological or theoretical support.  Thus, in the limit
$N_c\rightarrow \infty$, we believe that no tunnelling is supported
and the domain walls become stable.  In Section~\ref{sec:Nc3} we shall
extrapolate these results to the $N_c = 3$ limit and, although we lose
the theoretical control gained in the large $N_c$ limit, we argue that
the qualitative picture might remain the same.

\subsection{Heavy degrees of freedom}
\label{sec:heavy-degr-freed}

As discussed in Section \ref{sec:heavy}, we cannot properly discuss
tunnelling with Equation~(\ref{eq:Leff}) where the gluon degrees of
freedom are integrated out and replaced by their vacuum expectation
values: such a theory cannot describe the strings which are
responsible for the domain wall decay described in
Section~\ref{sec:estimates}.  Instead, we must take one step back and
describe the dynamics of the gluon condensate by considering the
original Lagrangian~\cite{Halperin:1998rc,Halperin:1998bs} which
includes the complex gluon degrees of freedom $h$.  In this effective
theory, the complex gluonic field will sit in a Mexican-hat potential
with $|\langle h\rangle|=$ const.  The string represents a localized
region about which $h = 0$.  In the effective theory~(\ref{eq:Leff})
this represents a singularity, but here the singularity is allowed,
although it is energetically costly.  The string tension is associated
with the energy cost required to form the string and this is directly
related to the height of the peak of the Mexican-hat potential.

The effective potential of the original
Lagrangian~\cite{Halperin:1998rc,Halperin:1998bs} which includes the
complex gluonic degrees of freedom $h$ is given by:\footnote{Note: the
  expression~(\ref{d6}) is valid only for one branch.  For a more
  accurate expression and treatment of the subsequent minimisation
  process which carefully accounts for the different branches, see the
  original paper~\cite{Fugleberg:1998kk}.}
\wide{
  \begin{equation}
    \label{d6}
    V (\theta, h,U) = \left(\frac{h}{4N_c} \log \left[ \left( \frac{h}{2 e
            E} \right)^{N_c} \frac{ {\rm Det} \, U }{ e^{-i\theta} } \right] -
      \frac{1}{2} {\rm Tr} \, M U\right) + \text{H.c.}
  \end{equation}
}{
\begin{equation}
  \label{d6}
  V (\theta, h,U)  =
  \left(\frac{h}{4N_c} \log \left[ \left( 
        \frac{h}{2 e E} \right)^{N_c}
      \frac{ {\rm Det} \, U }{ e^{-i\theta} } \right] -  \tfrac{1}{2}
    {\rm Tr} \, M U\right) + \text{H.c.}
\end{equation}
}
This satisfies all the conformal and chiral anomalous Ward identities,
has the correct large $N_c$ behaviour etc.\
(see~\cite{Halperin:1998rc,Halperin:1998bs,Fugleberg:1998kk} for
details).  Integrating out the heavy $h$ field will bring us back to
Equation~(\ref{eq:Leff}). (It is interesting to note that the
structure of Equation~(\ref{d6}) is quite similar in structure to the
effective potential for SQCD~\cite{Taylor:1983bp} and
gluodynamics~\cite{Migdal:1982jp}.)

Let us stop for a moment to consider the form of this potential.  When
we integrate out the heavy gluon degrees of freedom, we replace $h$
with its vacuum expectation value $\langle h \rangle$. The vacuum
expectation value of the $h$ is given by
\begin{equation}
\langle h \rangle = 2E\exp\bigl(-i\sum\phi_i/N_c\bigr) = 2E
e^{-i\phi_S/N_c}
\end{equation}
such that~(\ref{d6}) becomes
\begin{equation} 
\label{d7}
V(U)\Big|_{\genfrac{}{}{0pt}{1}{\!\!\!\!\!\theta=0}{h=\langle
h\rangle}} = - E \cos \frac{\sum \phi_i}{N_c} - \sum M_i \cos \phi_i\,,
\end{equation}
in agreement with~(\ref{7}).  With $h$ fixed at its vacuum expectation
value, the singlet combination $\phi_S=\sum{\phi_i}$ exhibits the
$\UU(1)$ topology described above and our domain walls are stable.
Now, however, we allow the gluon condensate to fluctuate.  We
parameterize these fluctuations in polar coordinates by a radial
component $\rho$, and an angular component $\zeta$:
\begin{equation}
\label{eq:heavy-fields}
h=\rho e^{i\phi_\zeta}\langle h \rangle = \rho
\exp\left(i\frac{\zeta}{f_\zeta}\right)\langle h \rangle\,.
\end{equation}
Here $f_\zeta$ is the decay constant and $\phi_\zeta$ is the
dimensionless phase angle.  The fields $\rho$ and $\zeta$ are both
heavy, real, physical fields.  In the large $N_c$ limit, their masses
are much heavier than the pion or $\eta'$ masses: $m_\zeta \sim m_\rho
\gg m_{\eta'}$.  The role of the $\zeta$ field will be important in
explaining the physics of the cusp singularities in~(\ref{eq:Leff})
and will be discussed in Section~\ref{sec:cusps}.  For now we set
$\theta$ and $\phi_\zeta$ to zero.  In terms of the remaining physical
fields, the potential $V$ becomes
\begin{equation}
\label{eq:Vhreal}
V(\rho,\phi_S)= +E\rho\bigl(\log\rho-1\bigr)\cos\frac{\phi_S}{N_c}\,,
\end{equation}
where we have neglected the terms proportional to $M$ since they only
contribute a constant offset due to the $\phi_T$ field.  For an early
phenomenological discussion of the potential~(\ref{eq:Vhreal}) without
the $\phi_S$ fields, see~\cite{Gomm:1986ut}.

Now the combined degrees of freedom $\rho$ and $\phi_S$ are no longer
restricted to the circle $\rho=1$ as they were in the effective
theory~(\ref{eq:Leff}).  The $\UU(1)$ topology is no longer a
constraint of the fields and thus, the walls are not topologically
stable.  Instead, the restriction of $\rho\approx 1$ is dynamical, and
made by a barrier at $h=0$.  Thus, with these degrees of freedom, the
fields parameterize the plane, \pagebreak[3] however, the potential is
that of a tilted Mexican-hat with a barrier at $h=0$.  The barrier is
high enough that a domain wall interpolating around the trough of the
hat is classically stable.  If the barrier were infinitely high as we
assumed when we fixed $\rho=1$ as we did in~(\ref{eq:Leff}), then the
$\phi_S$ field could wind around the barrier and would be
topologically stable.  With a finite barrier, however, the field can
tunnel through the barrier as described above.  This situation is
analogous to the case of the string and peg shown in
figure~\ref{fig:3dhomo}.  We show a more accurate picture\footnote{We
are indebted to Misha Stephanov for suggesting this nice intuitive
picture for explaining the domain wall decay mechanism.} of the
barrier~(\ref{eq:Vhreal}) in figure~\ref{fig:hat1}.  The relative
heights of the the peak and troughs are given below:
\begin{eqnarray}
\Delta V_{\rm Peak} &=& \Delta V_1 +\Delta V_2 = E \sim N_c^2\,, 
\label{eq:DV}\\
\Delta V_1 &=& E \cos\frac{\pi}{N_c}\sim N_c^2\,, 
\label{eq:DV1}\\
\Delta V_2 &=& E \left(1-\cos\frac{\pi}{N_c}\right)\sim 1\,. 
\label{eq:DV2}
\end{eqnarray}
We have emphasised the $N_c$ dependence here because the effective
theories are really only well justified in the large $N_c$ limit.

\FIGURE[ht]{\begin{centering}
\psfrag{BB}{$\Delta V_1\sim N_c^2$}
\psfrag{CC}{$\Delta V_2\sim 1$}
\psfrag{phi1}{$\phi_S=\pi$}
\psfrag{phi2}{$\phi_S=0$}
\includegraphics[width=\textwidth]{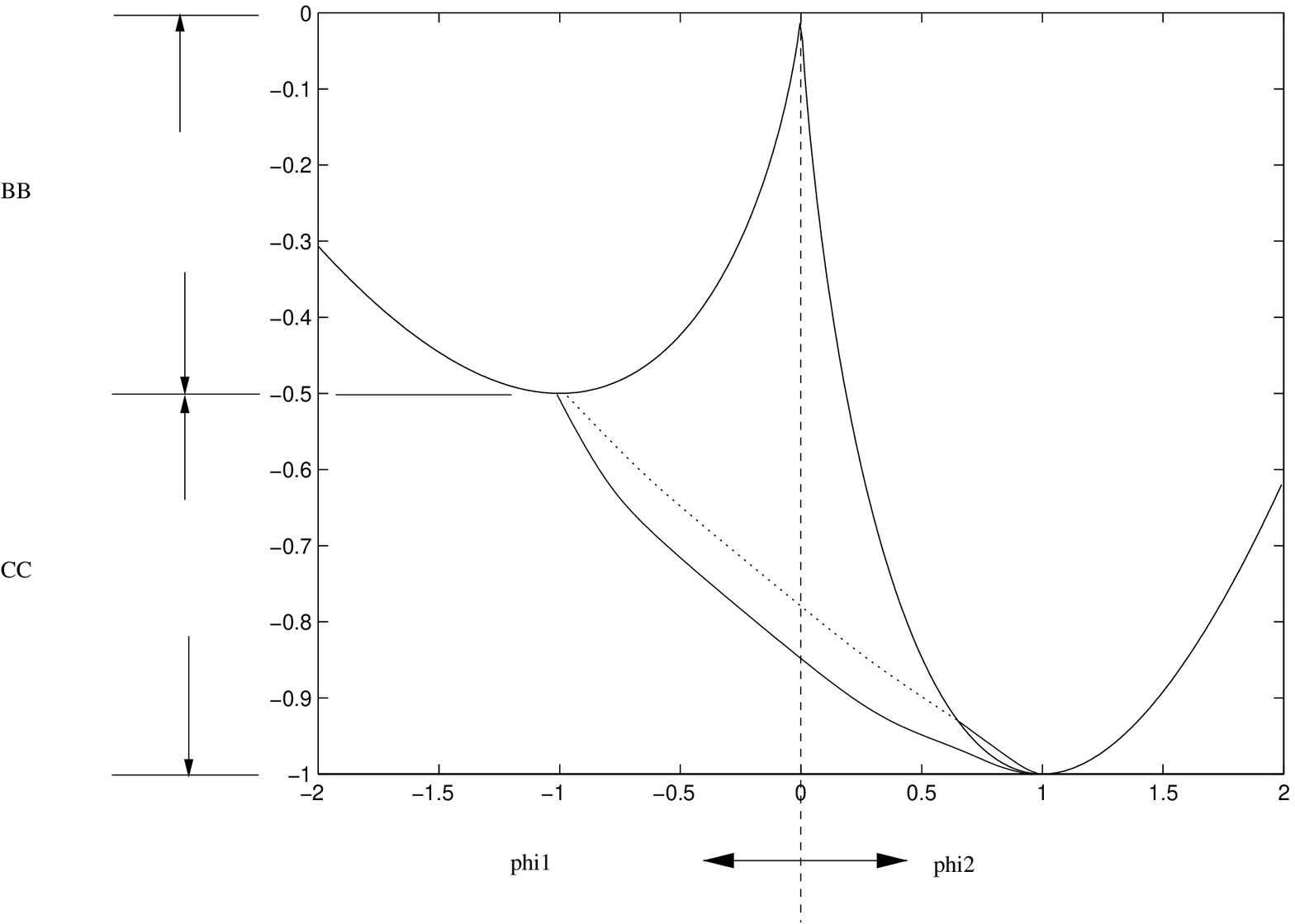}
\caption{Profile of the ``Mexican-hat'' potential~(\ref{eq:Vhreal}).
The slice is made along the axis through $\phi_S=\pi$ to the left and
$\phi_S=0$ to the right.  The trough of the potential lowers from the
cusp at $\rho=1$, $\phi=\pi$ where $V=-E\cos(\pi/N_c)$ down to the
vacuum state $V_{\rm min}=-E$. The hump where $V=0$ is at the origin
is where $h=\rho\exp(i\phi_S/N_c)=0$ and hence the singlet field
$\phi_S$ can have any value at this point.  It is by passing across
this point that a QCD domain wall can tunnel and a hole can form.  The
important qualitative features of this potential are the height of the
central peak $\Delta V_{\rm Peak}$ (\ref{eq:DV}), the relative heights
of the cusp to the peak $\Delta V_1$ (\ref{eq:DV1}) and the trough to
the cusp $\Delta V_2$ (\ref{eq:DV2}). \label{fig:hat1}}
\end{centering}}

\looseness=1The most important property of the potential is the following: The
absolute minimum of the potential in the chiral limit corresponds to
the value $V_{\rm min}=-E$ which is the ground state of our world with
$\rho=1$ and $\phi_s=0$.  At the same time, the maximum of the
potential~(\ref{eq:Leff}), where one branch changes to another one is
$V=-E\cos(\pi/N_c)$ where $\phi_S = \pi$ (we are still taking
$\theta=0$).  This corresponds to the point $\rho=1$ and $\phi_S=\pi$
in the potential~(\ref{eq:Vhreal}).  Thus, the trough of the
Mexican-hat is given where $h=\langle h \rangle$, i.e. at radius
$\rho=1$ and the maximum $V=-E\cos(\pi/N_c)$ of the
potential~(\ref{eq:Leff}) is exactly the barrier through which the
$\eta'$ field interpolates to form the QCD domain wall. It is
important to note that the height of the barrier for the
potential~(\ref{eq:Leff}) is numerically is quite high $\sim E
(1-\cos\frac{\pi}{N_c})$, but vanishes in large $N_c$ limit.  Indeed,
in this limit, the peak of the barrier is degenerate with the absolute
minimum $V_{\rm min}=-E$ of the potential as it should
be.\footnote{Remember, the $\eta'$ direction becomes flat in the large
$N_c$ limit as $m_{\eta'}\rightarrow 0$.}  The height of this barrier
describes how much the Mexican-hat is tilted.

The other important property of the potential~(\ref{eq:Vhreal}) is its
value where the singlet phase $\sum \phi_{i}$ is not well defined at
the peak of the Mexican-hat.  From~(\ref{eq:Vhreal}) it is clear that
this occurs for $h=0$.  It is this peak of the Mexican-hat in
figure~\ref{fig:hat1} (or the peg in figure \ref{fig:3dhomo}) that
classically prevents $\rho\rightarrow 0$ and that makes the QCD domain
wall classically stable.  When a hole tunnels through the wall, it
must be surrounded by a ``string'' where the field passes through the
region $h=0$.  The height of the peak thus \pagebreak[3] contributes to the
energetic cost of creating such a string (and hence the hole).  The
potential~(\ref{eq:Vhreal}) vanishes at the centre of the string
$V(h=0)=0$, which implies that the barrier at $h=0$ is
quite high: $\Delta V_{\rm Peak} = E$.  As expected, the barrier at
$h=0$ should be order of $E\sim N_c^2$ in contrast with the barrier to
the $\eta'$ domain wall where one expects a suppression by some power
of $N_c$.  We also note that the total number of distinct classically
stable solutions can be estimated from the condition that $\cos(\pi
k/N_c)>0$ where the barrier for the $\eta'$ field is still lower than
the peak $h=0$.  Thus, $-N_c/2<k<N_c/2$ where $k$ labels the winding
number of the solution.  Thus we see that for $N_c\leq 2$ there is no
admissible classically stable solution, but for $N_c\geq 3$
classically stable solutions are allowed. In our model with
potential~(\ref{eq:Vhreal}) we see explicitly that, for $N_c=3$, there
is one classically stable domain wall while for $N_c=2$ there are no
classically stable solutions.
\vspace{2em}

\subsection{String and wall tensions}
\label{sec:string-wall-tensions}

In order to make the semi-classical estimates of the decay rate (see
Equations~(\ref{d3}) and~(\ref{d5})) we must estimate the string
tension $\alpha$ and the wall tension $\sigma$.  We start with the
string tension $\alpha$.

Within the effective theory~(\ref{d6}) true strings are not supported
because the $\UU(1)$ symmetry is broken by the anomaly.  In the large
$N_c$ limit, however, this symmetry is restored (and the $\eta'$
becomes massless).  Thus, it makes sense to consider global $\UU(1)$
strings in the large $N_c$ limit.  To properly estimate $\alpha$ at
finite $N_c$ one must numerically minimizes the energy of a
configuration with a domain wall bounded by a string-like
configuration.  This could be done numerically, but is beyond the
scope of the present paper.  In the large $N_c$ limit, the estimates
presented here become reliable.

To estimate the string tension $\alpha$, we first consider an isolated
global string in a flat Mexican-hat potential
\begin{equation}
\label{eq:scalarfield}
{\mathcal{L}}=\frac{1}{2}\partial_\mu \Phi^* \partial^\mu \Phi -
V(|\Phi|)\,,
\end{equation}
where $\Phi$ is a generic complex scalar field that will be identified
with the glue-ball field $h$ discussed in
Section~\ref{sec:heavy-degr-freed}.  To match $V(|\Phi|)$
with~(\ref{d6}) we must take the limit $N_c\rightarrow\infty$ so that
the $\UU(1)$ symmetry $\Phi\rightarrow e^{i\alpha}\Phi$ (equivalently
$h \rightarrow e^{i\alpha}h$) is restored.  In this case, a string
lying along the $z$-axis with winding number $n$ will be described by
the complex field configuration
\begin{equation}
\Phi(r,\theta) = \rho(r,\theta)e^{i\phi(r,\theta)}=\rho(r)e^{in\theta}
= \rho(r)\exp \left(in\tan^{-1}\frac{y}{x}\right)
\end{equation}
in the $x$-$y$ plane.  The exact radial dependence $\rho(r)$ will be
such as to minimize the energy density along the string and will
minimize the energy density
\begin{eqnarray}
\label{eq:tension}
\label{eq:tension1}
\alpha &=& \iint \left(\frac{1}{2}\partial_i\rho\partial_i\rho
+\frac{1}{2} \rho^2\partial_i\phi\partial_i\phi
+V(\rho)-V_{\mathrm{min}}\right) \mathrm{d}x\mathrm{d}y\,,
\\
\label{eq:tension2}
&=& 2\pi\int
\left(\frac{1}{2}\left(\frac{\mathrm{d}\rho}{\mathrm{d}r}\right)^2
+\frac{n^2\rho^2}{2r^2} +V(\rho)-V_{\mathrm{min}}\right) r\mathrm{d}r\,.
\end{eqnarray}
We have assumed here that the fields $\rho$ and $\phi$ are canonically
normalized.

Within some radius $r_s$, the radial dependence of the string will
vary from $0$ to $\langle\rho\rangle$ and there will be a core energy
contribution to the string tension
\begin{equation}
\alpha_0 = \pi r_s^2 V_0\,,
\end{equation}
where $V_0$ represents an average core energy.  Far away from the
string, $\rho$ will assume its vacuum expectation value and the total
string tension (energy density per unit length) will be
\begin{equation}
\label{eq:tension4}
\alpha \sim \alpha_0+ \pi\int_{r_s}^R
\frac{n^2\langle\rho\rangle^2}{r}\mathrm{d}r \sim
\alpha_{\mathrm{core}} +\pi n^2\langle\rho\rangle^2\log R\,,
\end{equation}
where we have absorbed all contributions from the region of size $r_s$
into the constant $\alpha_{\mathrm{core}}$.  For an isolated string,
the string tension is infinite, but in most cases, the lateral extent
of the string is limited by some upper radius $R$ that must be
determined by the dynamics of the strings.  In our case, the string is
embedded in the end of a domain wall, so the relevant scale for $R$
will be the wall thickness.\footnote{To justify the use
of~(\ref{eq:tension4}) the radius of the string $r_s$ must be much
smaller than the wall thickness.  We shall show that this is the case
in~(\ref{eq:string_radius}).}

Far from the core of the string, the only degree of freedom is the
dimensionless phase $\phi$:
\begin{equation}
\Phi=\langle\rho\rangle e^{i\phi}\,.
\end{equation}
Thus, the kinetic term is
\begin{equation}
\frac{1}{2}\langle\rho\rangle^2\partial_\mu\phi\partial^\mu\phi\,.
\end{equation}
The normalization of the field $\rho$ is such that this term reduces
to the appropriate canonical kinetic term for the appropriate
Goldstone field described by Equation~(\ref{eq:scalarfield}).  In our
case, the string is composed of the $\eta'$ field and the relevant
phase $\phi$ interpolates from $0$ to $2\pi$ connecting the two sides
of the QCD domain wall.  For this $\eta'$ string, the the relevant
phase is $\phi=2\eta'/(f_{\eta'}\sqrt{N_f})$ is given in~(\ref{1}):
thus $\langle\rho\rangle^2=N_f f_{\eta'}^2/4$, and the tension is:
\begin{equation}
\label{eq:stringtension1}
\alpha \sim \alpha_{\mathrm{core}} +\frac{\pi}{4} N_f
f_{\eta'}^2n^2\log R \sim \alpha \sim \alpha_{\mathrm{core}}
+\frac{\pi}{2} f_{\pi}^2\log R\,,
\end{equation}
where we have set $N_f=2$ and $n=1$ (single winding) in the last
equation, as we have been doing.

To determine $\alpha_{\mathrm{core}}$, one must actually minimize the
string tension.  This requires full knowledge of the potential
$V(\rho)$ and the nature of the field $\rho$.  We do not have this
information.  In general, however, we expect that the scale of the
core is set by the mass of the heavy field $\rho$
\begin{equation}
r_s\sim m_\rho^{-1} \sim 1\,,
\end{equation}
and thus, in the large $N_c$ limit, the string core size becomes much
smaller than the domain wall thickness~(\ref{eq:QCDwidth})
\begin{equation}
\label{eq:string_radius}
r_s \sim m_\rho^{-1} \sim 1 \ll \mu^{-1} \sim m_{\eta'}^{-1} \sim
\sqrt{N_c}\,,
\end{equation}
so one can think of the string as embedded in the edge of the domain
wall.  In this environment, the outer radius of the string $R$ is the
same order as the thickness of the domain wall.  This justifies the
approximation of the string tension~(\ref{eq:stringtension1}) as that
of a free string with outer radius $R$: if the string core had been of
comparable size to the domain wall, then this approximation would be
inaccurate, and we would have had to minimize the energy of the
combined system of a domain wall bounded by a string.  This situation
is much more difficult to solve because the ``string'' would no longer
have cylindrical symmetry.

Unlike the tension of a global string, which has a logarithmic
contribution from the bulk~(\ref{eq:stringtension1}), the domain wall tension
(c.f.\ Equation~(\ref{tension}) come exclusively from the core dynamics.
Due to the quadratic kinetic terms, there is an equipartition of
energy and the tension is essentially
\begin{equation}
\label{eq:estimate_wall_tension}
\sigma \sim \Delta z V_0 \,,
\end{equation}
where $\Delta z$ is the wall thickness and $V_0 \sim V_{\mathrm{core}}
- V_{\mathrm{min}}$ is the average potential energy near the core.
For most domain walls, the thickness is governed by the mass of the
relevant field.  Thus, for QCD domain wall tension, we have the
following contributions from the pion and $\eta'$ transitions
respectively (see potential~(\ref{eq:Vfull})):
\begin{eqnarray}
\label{eq:tension_estimates}
\label{eq:tension_pi}
\sigma_{\pi}&\sim& m_{\pi}^{-1} 2M \sim m_{\pi}^{-1}
m_q|\langle\bar{\Psi}\Psi\rangle| \sim f_\pi^2 m_\pi\,,
\\
\label{eq:tension_eta}
\sigma_{\eta'}&\sim& m_{\eta'}^{-1}E\left(1-\cos\frac{\pi}{N_c}\right)
\sim \frac{E}{N_c^2 m_{\eta'}}\,,
\end{eqnarray}
which agree qualitatively with the precise calculations~(\ref{a3}).
Thus, in the chiral limit $m_q\rightarrow 0$, only the $\eta'$
contribution remains relevant.  Numerically, even with finite quark
masses, the $\eta'$ contribution is much larger than the pion
contribution.  There is another contribution, however, due to heavy
degrees of freedom which we estimate in the next section.

\subsection{Cusps}
\label{sec:cusps}

As was emphasized in~\cite{Kogan:1998dt} for supersymmetric QCD, the
presence of a cusp singularity in the effective
potential~(\ref{eq:Leff}) can indicate that heavy degrees of freedom
are playing a role in the physics.  As far as the low energy degrees
of freedom are concerned, (the pions and $\eta'$), the effective
potential~(\ref{eq:Leff}) is valid on the scales associated with these
degrees of freedom.  If we were to properly include the heavy degrees
of freedom, we would find that the cusps are actually smooth, but only
on scales comparable to the mass of the heavy particles $m_h \gg
m_\pi$.  In the centre of the domain wall, where the potential has the
cusp, what is really happening is that the heavy fields are making a
rapid transition with a scale length of $\Delta z\sim m_h^{-1}$.

Analogously to~\cite{Kogan:1998dt}, the contribution of this heavy
transition to the domain wall tension can be estimated
by~(\ref{eq:estimate_wall_tension}) using the heavy fields introduced
in~(\ref{eq:heavy-fields}).  The mass scale $m_\zeta$ sets the width
for the transition, but to determine the scale for the potential $V_0$
requires more work.  One can estimate this from the mass term of the
$\zeta$ field as follows.

In order to correctly allow the $\eta'$ field to interpolate from
$\theta=0$ to $\theta=2\pi$, the field $\zeta$ enters the picture as a
phase in the following combination with the $\eta'$ and $\theta$
angle:
\begin{equation}
\label{eq:4}
\exp\left(i\frac{\theta}{N_c} +i\frac{\eta'}{\sqrt{2}f_{\eta'}N_c}
+i\frac{\zeta}{f_\zeta}\right) =
\exp\left(i\frac{\theta+N_c\zeta/f_\zeta}{N_c}
+i\frac{\eta'}{\sqrt{2}f_{\eta'}N_c}\right).
\end{equation}
In order to allow the $\eta'$ field to switch from one branch of the
potential to another, $\zeta$ must shift in such a way that $\theta
\rightarrow \theta+2\pi$.  Thus, with our definition, $\zeta$ makes
the transition from $0$ to $2\pi f_\zeta/N_c$.  Since there is an
equipartition between the kinetic and potential energies, we can
estimate the contribution to the wall tension from the $\zeta$ field
by using the kinetic term:
\begin{equation}
\label{eq:tension_heavy}
\sigma_\zeta \sim \Delta z \left(\frac{\Delta\zeta}{\Delta z}\right)^2
\sim \frac{4\pi^2 f_\zeta^2}{N_c^2\Delta z} \sim \frac{m_\zeta
f_\zeta^2}{N_c^2}.
\end{equation}
The numerical contribution of this term to the wall tension might be
quite large since it is proportional to the heavy mass $m_\zeta$.
Expanding~(\ref{eq:4}) and noting that the potential will depend on
some energy scale $\sim N_c^2$ related to gluonic physics, we estimate
the mass term as
\begin{equation}
m_\zeta^2 \sim N_c^2/f_\zeta^2\,.
\end{equation}
Using the standard assumption that masses do not depend on $N_c$, we
have that $m_\zeta \sim 1$.  Thus, $f_\zeta \sim N_c$ and
\begin{equation}
\label{eq:zeta_nc}
\sigma_\zeta \sim 1\,.
\end{equation}

Thus, the presence of cusps in an effective theory signals that heavy
degrees of freedom might play an important quantitative role through
contributions like $\sigma_\zeta$, in agreement with the conclusion
reached in~\cite{Kogan:1998dt}.  If we make the same estimates for the
$\eta'$ meason, we get $\sigma_{\eta'}\sim\sqrt{N_c}$ instead
of~(\ref{eq:zeta_nc}) due to the unique way that the $\eta'$ couples
to the $\theta$ parameter through the term
$(\theta-i\log\textrm{Det}~U)$, and the unique $N_c$ dependence:
$m_{\eta'}\sim N_c^{-1/2}$.

There is one other cusp to consider in~(\ref{eq:Vhreal}) and
figure~\ref{fig:hat1} where $\rho=0$.  This cusp will somehow be
smoothed on small scales affecting short distance physics.  This will
definitely alter the core energy of the
string~(\ref{eq:stringtension1}).  It will not, however, affect the
long distance behaviour of the string, which is typically the most
important contribution and which is ultimately the source of the large
$N_c$ dependence.

\subsection{Large $N_c$ limit: summary}
\label{sec:large-n_c-limit}

As we have suggested earlier, unless there are peculiar degrees of
freedom that greatly increase the domain wall tension, all of our
results come under theoretical control in the large $N_c$ limit
$N_c\rightarrow\infty$.  The motivation for this comes from
figure~\ref{fig:hat1}.  In the large $N_c$ limit, the central barrier
becomes extremely high as it is of order $N_c^2$ while the trough
becomes flat.  Thus, we approach the picture of
figure~\ref{fig:2dhomo} where the probability of nucleation becomes
zero.  In this limit, the domain walls become stable.  To demonstrate
this, we must show that, in this limit, the probability of creating a
hole in the wall~(\ref{d3}) falls to zero by verifying the
assertions~(\ref{eq:tensions_Nc}) and~(\ref{eq:decay_rate_Nc}).

First, consider the $N_c$ dependence of the string tension
$\alpha$~(\ref{eq:stringtension1}):
\begin{equation}
\alpha \sim \alpha_{\mathrm{core}} + \frac{\pi}{2}f_\pi^2\log
m_{\eta'}^{-1}\,.
\end{equation}
The last term provides an $N_c$ dependence of at least $N_c\log N_c$.
The core term can possible increase the tension, however, as we wish
to demonstrated that the decay rate is suppressed, we must assume the
worst possible case and take the lowest possible bound for $\alpha$.
Thus, we consider only the last term and set
\begin{equation}
N_c \lesssim N_c \log N_c \lesssim \alpha\,.
\end{equation}
This bound is quite robust.

Now consider the three contributions to the domain wall
tensions (\ref{eq:tension_estimates}) and (\ref{eq:tension_heavy})
\begin{eqnarray}
\label{eq:1}
\sigma_{\pi}&\sim& f_\pi^2 m_\pi \sim N_c \sqrt{m_q}\,,
\\
\label{eq:2}
\sigma_{\eta'}&\sim& m_{\eta'}^{-1}E/N_c^2 \sim \sqrt{N_c}\,,
\\
\label{eq:3}
\sigma_\zeta &\sim& m_\zeta f_\zeta^2/N_c^2 \sim 1\,.
\end{eqnarray}
To preserve the qualitative effects of the chiral limit, we first take
this limit.  Thus, despite of the fact that the $\pi$ meson cloud is
of a much larger size than the $\eta'$ transition, its contribution to
$\sigma$ is much smaller $m_\pi\sim \sqrt{m_q} \rightarrow 0$ in both
the large, and the physical $N_c$ limits.  Furthermore, it is a widely
accepted assumption that, in QCD, all physical masses are of order $1$
in the large $N_c$ limit.  The $\eta'$ mass is of course an
exception~(\ref{eq:metaNc}), but only in the chiral limit as we have
taken here: at fixed $m_q$, even $m_{\eta'}\sim1$ in the large $N_c$
limit.  The heavy degree of freedom $\zeta$, however, is related to
gluonic physics, and should thus not be sensitive to the chiral limit.
If this is the case, then $m_\zeta \sim 1$ and the $\eta'$
contribution dominates in the large $N_c$ limit.

The reason that we have drawn so much attention to the point that in
QCD, all masses seem to be of order $1$ is that similar domain walls
were discussed in the context of supersymmetric
QCD~\cite{Kogan:1998dt}.  There, a similar analysis to that performed
in Section~\ref{sec:QCDwalls} resulted in a domain wall of tension
$\sigma\sim 1 \ll N_c$, analogous to our estimate~(\ref{eq:2}).
However, in supersymmetric theories, this is in direct contrast with
the analytic BPS lower bound of order $N_c$.
In~\cite{Gabadadze:1999pp}, it was subsequently conjectured that heavy
degrees of freedom at the cusp give a contribution similar
to~(\ref{eq:3}) but that, in order to reconcile this result with the
BPS limit, the relevant mass of the heavy degrees of freedom has a
peculiar dependence on the number of colours:~\mbox{$m_\zeta \sim N_c$}.

In QCD, there is no analogue of the BPS bound which can guide us, and
we are not convinced that any such degrees of freedom exist with
$m_\zeta \sim N_c$.  However, imagining that they might exist, we take
as a worst case\footnote{Notice that, if we do not first take the
chiral limit, then the pion contribution to the domain wall provides a
$\sigma_\pi \sim N_c$ contribution that will dominate in the large
$N_c$ limit.  Such behaviour is not related to some mass $m_\pi \sim
N_c$.  Thus, the pion contribution gives $\sigma\sim N_c$ behaviour
when $m_q\neq 0$ is held fixed.  Might it be possible that the
resolution to the problem of the BPS bound in the supersymmetric case
lies in an analogous degree of freedom to the QCD pion field which is
not directly connected to the $\UU(1)$ anomaly?  These degrees of
freedom might have been neglected in the relevant effective theory,
yet may supply the required $N_c$ dependence without requiring a field
with strange $N_c$ mass dependence.  While we cannot say more about
this here, the possibility supports the widely believed assumption
that all masses in QCD are of order $1$ in the large $N_c$ limit.}
\begin{equation}
\sqrt{N_c} \lesssim \sigma \lesssim N_c\,.
\end{equation}

Combining the results of Section~\ref{sec:estimates}, we have the
following $N_c$ dependences
\begin{subequations}
\label{eq:Nc-Results}
\begin{align}
\alpha &\sim N_c\log N_c\,,&
\sigma &\sim \sqrt{N_c}\,,\\
R_c &\sim \sqrt{N_c}\log N_c\,,&
\mu^{-1} &\sim \sqrt{N_c}\,,\\ 
S_0 &\sim N_c^2\,,&
P &\sim e^{-N_c^2}\,.
\end{align}
\end{subequations}
Notice also that, although both the critical radius for nucleation
$R_c$ and the wall thickness $\mu^{-1}$ increase, the ratio $\mu R_c
\sim \log N_c \gg 1$ and the semiclassical approximation used to
derive the decay rate~(\ref{d3}) becomes justified in the large $N_c$
limit.

Allowing for the possibility that the contribution $\sigma_\zeta$ of
the heavy field has an $N_c$ dependence as conjectured in
supersymmetric QCD, the tunnelling rate should still be exponentially
suppressed:
\begin{subequations}
\begin{align}
\alpha &\gtrsim N_c \log N_c,&
\sigma &\sim N_c, \\
R_c &\gtrsim \log N_c,&
\mu^{-1} &\sim \sqrt{N_c},\\ 
S_0 &\gtrsim N_c,&
P &\lesssim e^{-N_c}.
\end{align}
\end{subequations}
In this case, our use of the semi-classical approximation --- with
independent calculations of the string and domain wall tensions --- is
no longer justified, however, the exponential contribution will still
remain.  The only way to ruin the qualitative picture would be to
argue that a minimal contribution to the wall tension $\sigma$ exists
that is $\gtrsim N_c^{3/2}$ which is required to support the $\eta'$
transition $\Delta\theta = 2\pi$.  At present, we see no evidence or
justification for such a contribution.

\subsection{$N_c=3$}
\label{sec:Nc3}

We have argued that, in the large $N_c$ limit while preserving
approximate chiral symmetry, QCD domain walls are stable to leading
order and that, to leading order, they can decay through a tunnelling
mechanism.  We now provide some  estimates of the magnitude
of these effects extrapolating back to the realistic limit of $N_c=3$.
In this limit, we assume that the most relevant contribution to the
domain wall tension is from the $\eta'$: the pion contribution is
suppressed by the ratio $m_\pi/m_{\eta'}$ and numerically contributes
only $10\%$ of the tension.  Neglecting the contribution from the
heavy field $\zeta$ is more difficult to justify: numerically, both
this and the $\eta'$ contributions are likely important, but it is
reasonable to neglect them for an order of magnitude estimate.  Thus
we take
\begin{equation}
\label{d1}
\sigma = \frac{4N_cf_{\pi}\sqrt{E}}{\sqrt{2}} \left(1-
\cos\frac{\pi}{2N_c}\right).
\end{equation}
In this formula, the $\sqrt{2}$ in denominator should be replaced by
$\sqrt{N_f}$ for an arbitrary $N_f$; however in all numerical
estimates, we shall use $N_f=2$ which we believe is very good
approximation in the limit $m_u\simeq m_d \ll m_s$ as equations
(\ref{9}) and (\ref{10}) suggest.  Besides that, for $N_c\geq 3$, one
can approximate $1-\cos[\pi/(2N_c)]\simeq \pi^2/(8N_c^2)$ such that
Equation~(\ref{d1}) takes the simple form
\begin{equation}
\label{d2}
\sigma=
\frac{\pi^2f_{\pi}\sqrt{E}}{2\sqrt{2}N_c}
\end{equation}
which will be used for our numerical estimates.

As an estimate for the string tension, we make the following estimate
based on dimensional grounds:
\begin{equation}
\label{eq:stringtension}
\alpha \sim \sqrt{2E}\,.
\end{equation}
This estimate has the same $N_c$ dependence
as~(\ref{eq:stringtension1}).\footnote{The magnitude for $\alpha\sim
\sqrt{2E} \sim (0.28 {\rm GeV})^2$ should not be considered as a
strong overestimation.  Indeed, if one considers the $\pi$ meson
string ~\cite{Brandenberger:1998ew} which should be much softer (and
therefore, would possess much smaller $\alpha_{\pi}$) one finds,
nevertheless, that $\alpha_{\pi}\simeq \pi f_{\pi}^2$ is very close
numerically to this estimate for the $\eta'$ string tension.}

For numerical estimates, we set $N_c=3$, $N_f=2$, and use:
\begin{subequations}
\label{eq:numbers}
\begin{align}
\label{d8a}
\alpha & \simeq \sqrt{2E}\sim (0.29 \text{ GeV})^2 ,\\
\sigma & \simeq \frac{\pi^2 f_\pi \sqrt{E}}{2\sqrt{2} N_c} \sim
f_\pi(0.26\text{ GeV})^2 \sim (0.21 \text{ GeV})^3\\
R_c &=\frac{2\alpha}{\sigma}\sim \frac{2.4}{f_{\pi}}, \\
S_0 &=\frac{16\pi\alpha^3}{3\sigma^2} \sim \frac{256
N_c^2\sqrt{2E}}{3\pi^3f_{\pi}^2}\sim 130,
\label{d8d}
\end{align}
\end{subequations}
Although the estimate~(\ref{d8d}) must be treated as very rough, it is
nevertheless quite remarkable: In spite of the fact that all
parameters in our problem are of order $\Lambda_{{\rm QCD}}$, the
classical action $S_0$ may still be numerically large, and thus the
corresponding tunnelling probability~(\ref{d3}) might be quite
small. We do not see any simple explanation for this phenomenon except
for the fact that expressions~(\ref{d4}) and~(\ref{d8d}) for $S_0$
contains a huge numerical factor $16\pi/3$ of purely geometrical
origin.  In addition, since $S_0\sim N_c^2$, one expects an additional
enhancement in $S_0$, even for $N_c=3$.

At this point we must address the validity of extrapolating from the
large $N_c$ approximation to the physical region of $N_c=3$.
Qualitatively we expect the physics to be the same --- the
potential~(\ref{d6}) displayed in figure~\ref{fig:hat1} has the same
qualitative form for $N_c>2$: namely, the central peak is much higher
than the troughs and classically stable solutions are admitted.  As
long as $\Delta V_1 > 0$, the domain wall solutions remain classically
stable, susceptible only to nucleation by string: the qualitative
picture thus remains the same.  Including heavier degrees of freedom,
while certainly affecting the numerical results, should not modify
this qualitative picture.

Perhaps we should have included the effects of other light degrees of
freedom such as the other pions, kaons etc.  In the $N_c=3$ case,
these degrees of freedom are certainly relevant in general, but are
not related to the physics which governs the $\UU(1)_A$ QCD domain
walls.  We emphasize: QCD domain walls appear due to the discrete
symmetry $\theta\rightarrow\theta+2\pi$ and the way that the singlet
$\eta'$ meason uniquely interacts with the parameter $\theta$ through
the term $(\theta-i\log\mathrm{Det}~U)$.  Thus, to satisfy the given
boundary conditions, one expects the existence of a classical
configuration for the $\eta'$ field.  The other light degrees of
freedom exist as fluctuations in this classical background.
Numerically one should account for these light particles scattering
off of the domain walls, but for the qualitative estimates presented
here, these light mesons are inconsequential.  For the same reason, we
do not expect the other mesons ($\rho$, $\omega$ etc.) to change the
qualitative physical behaviour. The key point is the Mexican-hat shape
of the potential~(\ref{d6}) and that the potential of the central peak
--- where $\rho=0$ and the phase of the $\eta'$ phase is not
well-defined --- is much higher than any point in the trough defining
the low-energy theory (\ref{eq:Leff}).  This qualitative picture holds
in both the large $N_c$ and $N_c=3$ limits.  Low-energy fluctuations
will not affect the height of the peak which makes the $\eta'$ domain
walls classically stable.

The $\eta'$ mass serves to emphasize the validity of the large $N_c$
limit in another way: The approximation of the chiral limit~(\ref{14})
remains valid up until $N_c\sim 10$.  Thus, the physical case of
$N_c=3$ should be related to the chiral large $N_c$ limit that we have
explored in this paper.

One other interesting note: As we mentioned earlier, in QCD there is
only one dimensional parameter, $\Lambda_{{\rm QCD}}$, and it is thus
generally believed that the semi-classical approximation in QCD cannot
be parametrically justified.  Nevertheless, $R_c$ is numerically quite
large --- much larger than the width $\mu^{-1}$ of the domain
wall~(\ref{eq:QCDwidth}) which is set by $m_{\eta'}^{-1}$.  Therefore,
the semi-classical approximation~(\ref{d3}) is somewhat justified a
posteriori.

\subsection{Lifetime}

Now, we are prepared to make our last step and estimate the
probability of creating a hole with radius $R_c\sim
8N_c/(\pi^2f_{\pi})$:
\wide{
  \begin{equation}
    \frac{P}{T} \pi R_c^2
    \left[\sqrt{\frac{S_0}{2\pi}}\right]^{3}e^{-S_0}\times {\rm Det} \sim
    \pi R_c^2 E^{\frac{3}{4}} \left[\sqrt{\frac{S_0}{2\pi}}\right]^{3}
    e^{-S_0} \,.
    \label{d9}
  \end{equation}
}{
\begin{equation}
  \frac{P}{T} \pi R_c^2
  \left[\sqrt{\frac{S_0}{2\pi}}\right]^{3}e^{-S_0}\times {\rm Det} 
  \sim \pi R_c^2 E^{\frac{3}{4}} 
  \left[\sqrt{\frac{S_0}{2\pi}}\right]^{3} e^{-S_0} .
  \label{d9}
\end{equation}
}
We have estimated ${\rm Det}\sim E^{\frac{3}{4}}$ dimensionally.
Using the numerical values for $R_c$ and $S_0\simeq 130$ given
in~(\ref{eq:numbers}) we arrive to the following final result:
\begin{equation}
\label{d10}
\frac{P}{T} \sim ~~ 10^3e^{-130} {\rm GeV}~~ \sim ~~10^{-30} s^{-1}\,.
\end{equation}
The most amazing result of the estimate~(\ref{d10}) is the
astonishingly small probability for the decay $P\sim 10^{-50}$ GeV
which one might na\"\i{}vely expect to be on the GeV level. This small
number leads to a very large life time for the domain walls, and
consequently, makes them relevant to cosmology at the QCD scale.  Of
course, our estimation of $S_0$ is not robust, and even small
variation of parameters may drastically change our
estimate~(\ref{d10}), making it much larger or much smaller.  The
point we wish to emphasize here is that it is at least possible for
QCD domain walls to live long enough to have non-trivial physical
effects.

First of all, let us estimate an average size $l$ of a domain wall
before it collapses with an average lifetime of $\tau_l\sim l/c$.
This corresponds the situation when the probability of the decay is
close to one.  If $L$ is the Hubble size scale, $L\sim 30$~km, then
\wide{
  \begin{eqnarray}
    P &\simeq& \frac{P}{T}\cdot\frac{l^2}{\pi R_c^2}\cdot\tau_l\sim
    \frac{P}{T}\cdot\frac{L^2}{\pi
      R_c^2}\cdot\frac{L}{c}\cdot\left(\frac{l}{L}\right)^3 \sim 10^{-30}
    s^{-1} \cdot 10^{36}\cdot 10^{-4}s\cdot \left(\frac{l}{L}\right)^3
    \nonumber\\
    &\sim& 10^2 \left(\frac{l}{L}\right)^3 \simeq 1\,.
    \label{d11}
  \end{eqnarray}
}{
\begin{align}
  \nonumber
  P&\simeq \frac{P}{T}\cdot\frac{l^2}{\pi R_c^2}\cdot\tau_l\sim 
  \frac{P}{T}\cdot\frac{L^2}{\pi
    R_c^2}\cdot\tfrac{L}{c}\cdot\left(\tfrac{l}{L}\right)^3
  \sim
  10^{-30} s^{-1}
  \cdot 10^{36}\cdot   10^{-4}s\cdot \left(\frac{l}{L}\right)^3\\
  &\sim 10^2 \left(\frac{l}{L}\right)^3 \simeq 1.
  \label{d11}
\end{align}
}
Formula~(\ref{d11}) implies that the average size $l$ of the domain
wall before it collapses could be as large as the Hubble size $l\simeq
10^{-2/3}\cdot L\sim 0.2 L$.  It also implies that on a Hubble scale
domain wall $L$, the average number of holes which will be formed is
approximately $\langle n \rangle \sim (L/l)^2\sim 25$.  Finally, the
lifetime of a Hubble size domain wall is expected to be on the scale
of
\begin{equation}
\label{d12}
\tau_{l}\sim \frac{L}{\sqrt{\langle n \rangle}c}\sim 2\cdot 10^{-5}s\,,
\end{equation}
which is macroscopically large!

It is clear that all these phenomena are due to the astonishingly
small number~(\ref{d10}) which makes the link between QCD and
cosmology feasible.  This small number is due to the tunnelling
process rather than some special fine-tuning arrangements.  As we
mentioned above, we have not made any adjustments to the
phenomenological parameters used in the estimates. Rather, we have
used the standard set of parameters introduced in
Equation~(\ref{eq:Leff}).  To conclude the discussion of~(\ref{d10})
we would like to remind the reader that similar miracles related to
tunnelling processes happens in physics quite often. For example, the
difference in lifetime for U$^{238}$ and Po$^{212}$ under $\alpha$
decay is on the order of $10^{20}$ in spite of the fact that the
``internal'' physics of these nuclei, and all internal scales are very
similar.

We should be careful to qualify this result.  First, it assumes that
the effective Lagrangian~(\ref{eq:Leff}) is a good description of the
low energy physics of QCD including the physics of the $\eta'$ field.
This assumption is well justified in the large $N_c$ limit.  The
theory is not quantitatively justified as $N_c\rightarrow 3$ as we
discussed in Section~\ref{sec:Nc3}.  Nevertheless, the qualitative
features of the potential responsible for producing the domain walls
persist for all $N_c \geq 3$ as shown in figure~\ref{fig:hat1} so we
expect the qualitative features to remain.  Second, it assumes that
the semi-classical calculation leading to~(\ref{d9}) is justified.
This is certainly not justified a-priori, but is somewhat justified
a-posteriori.  Third, we have assumed $T=0$ for this analysis.
Thermal fluctuations may considerably reduce the lifetime of these
walls. While this is a concern for cosmological domain walls, it
should not be a concern at RHIC where the quenched approximation
should be valid (see also footnote~\ref{foot:PT}). Finally, the
interaction of the domain wall with nucleons can drastically change
the numerical properties of the domain wall due to the strong
interaction of all relevant fields.  In particular, the domain walls
may become much more stable in the presence of nucleons, and even
account for the strong self-interacting dark matter discussed
in~\cite{Spergel:1999mh,Wandelt:2000ad}.  We hope to return to these
considerations in a separate publication.

\section{Conclusion and future directions}
\label{sec:conclusion}

The main results of this paper is expressed by the
formulae~(\ref{eq:QCDetasol}) and (\ref{eq:Nc-Results}) where a new
type of a quasi-stable QCD matter --- the QCD domain wall (which could
have a macroscopically large size!) --- is described. The qualitative
picture for the QCD domain walls is exact in the chiral large $N_c$
limit where the QCD domain walls are stable.  To leading order, the
walls decay as discussed, though the lifetime~(\ref{d10}) of the QCD
domain walls is potentially very large.  If this is the case, such
objects might play an important role on the evolution of the early
universe soon after the QCD phase transition.  As a by-product, we
also found a new type of axion domain wall ($a_{\eta'}$) with QCD
structure.  This may also have further cosmological applications if
the axions exist.  See~\cite{Forbes:2000gr} for a first attempt in
these directions.

Unfortunately, we do not yet have a good microscopic picture of QCD in
the low temperature, low density regime with small $N_c=3$, thus we
cannot further justify the effective theory~(\ref{eq:Leff}).  This
analysis, however, show that there is at least the hope of forming
macroscopically large domain walls with QCD scale.  Additional support
for these objects comes from the high density regime we have better
control of the physics.  In this case, QCD domain walls nearly
certainly exist~\cite{Son:2000fh}.  Furthermore, if one accepts a
conjecture on quark-hadron continuity at low temperature with respect
to variations in the chemical potential $\mu$~\cite{Schafer:1998ef},
then one can make the following argument: If for large $\mu$ domain
walls exist, but for low $\mu$ they do not, then there should be some
sort of phase transition as one lowers $\mu$.  Thus, the continuity
conjecture supports the existence of quasi-stable QCD domain walls at
lower densities, at least down to the densities of hypernuclear
matter.  Coupled with the fact that gluon and quark condensates do not
vary much as one moves to the low density limit, we suspects that the
qualitative picture holds even for zero density.

It would be very exciting if QCD domain walls can be studied at RHIC
(Relativistic Heavy Ion Collider)~\cite{Stephanov:2001}.  Such a
research would be the first experimental attempt to directly study the
fundamental properties of the QCD vacuum structure.  In this case the
temperature would fall rapidly due to the rapid expansion and we
expect that we can neglect thermal fluctuations as we have here.

In any case, QCD domain walls are a very interesting feature of QCD
and may live long enough to affect many areas of particle physics and
astrophysics.  We look forward to further study of these creatures!

\acknowledgments

This work is supported in part by the Natural Sciences and Engineering
Research Council of Canada.  A.Z. wishes to thank Misha Stephanov and
Edward Shuryak for valuable comments and very useful discussions
regarding the possibility of studying QCD domain walls at RHIC. He is
also grateful to Paul Stenhardt for explaining his
works~\cite{Spergel:1999mh,Wandelt:2000ad} and for discussions of
possibility of QCD domain walls as a strongly interacting dark matter
candidate and thanks G.~Dvali, G.~Gabadadze and M.~Shifman for useful
discussions.  We also wish to thank a referee whose report
considerably improved our understanding of QCD domain walls and, we
hope, improved our presentation.

\end{document}